\documentclass[onecolumn]{svjour3} 

\usepackage[american]{babel}
\usepackage[utf8]{inputenc}
\usepackage{ae,aecompl}
\usepackage[numbers]{natbib}

\usepackage{bm}
\usepackage{color}
\usepackage{booktabs}
\usepackage{graphicx}
\usepackage{subfigure}

\usepackage{amsmath} 
\usepackage{amssymb}
\usepackage{commath}
\usepackage{dsfont}

\newlength{\arrow}
\newcommand{\R}{\ensuremath{\mathbb{R}}}           

\renewcommand{\vec}[1]{\bm{\mbox{#1}}}

\newcommand{\mat}[1]{\bm{#1}}

\newcommand{\innerL}[2]{\left\langle #1, #2 \right\rangle_{\SS}}
\newcommand{\innerD}[2]{\left\langle #1, #2 \right\rangle_{\mathcal{D}}}

\newcommand{\normL}[1]{\norm{#1}_{\SS}}

\renewcommand{\SS}{\Theta}

\newcommand{\SSpt}{\theta}

\newcommand{\SF}{\mathcal{F}}

\newcommand{\PM}{P}

\newcommand{\PD}[1]{\PM_{#1}}

\newcommand{\randvar}[1]{#1}

\newcommand{\randproc}[1]{#1}

\newcommand{\pdf}[1]{p_{\tiny{#1}}}

\newcommand{\supp}[1]{\texttt{Supp}\,{#1}}

\newcommand{\expval}[1]{E \left\lbrace  #1 \right\rbrace }

\newcommand{\mean}[1]{\mu_{#1}}

\newcommand{\stddev}[1]{\sigma_{#1}}

\newcommand{\estim}[1]{\widehat{#1}}

\newcommand{\corr}[1]{\mathcal{K}_{#1}}

\journalname{J Braz. Soc. Mech. Sci. Eng.}

\begin{document}

\title{Non-intrusive polynomial chaos expansion for topology optimization using polygonal meshes}
\titlerunning{}

\author{Nilton Cuellar \and Anderson Pereira \and Ivan F. M. Menezes \and Americo~Cunha~Jr}
\authorrunning{N. Cuellar \and A. Pereira \and I. F. M. Menezes \and A. Cunha~Jr}

\institute{N. Cuellar \and A. Pereira \and I. F. M. Menezes \at
				Pontifical Catholic University of Rio de Janeiro -- PUC--Rio,
				Department of Mechanical Engineering,
				Rua Marqu\^{e}s de S\~{a}o Vicente, 225, Rio de Janeiro, 22453-900, RJ, Brazil\\
				\email{ncuellar@tecgraf.puc-rio.br}\\
				\email{anderson@tecgraf.puc-rio.br}\\
				\email{ivan@puc-rio.br}
              \and
              A. Cunha~Jr \at
              Universidade do Estado do Rio de Janeiro -- UERJ,
		      Nucleus of Modeling and Experimentation with Computers -- NUMERICO,
			  Rua S\~{a}o Francisco Xavier, 524, Rio de Janeiro, 20550-900, RJ, Brazil\\
              \email{americo@ime.uerj.br}
}

\date{Received: date / Accepted: date}

\maketitle

\begin{abstract}
This paper deals with the applications of stochastic spectral methods 
for structural topology optimization in the presence of uncertainties. 
A non-intru\-sive polynomial chaos expansion is integrated into a topology 
optimization algorithm to calculate 
low-order statistical moments of the mechanical-mathematical model response. 
This procedure, known as robust topology optimization, can optimize the mean of the 
compliance while simultaneously minimizing its standard deviation. 
In order to address possible variabilities in the loads applied to the
mechanical system of interest, magnitude and direction of the external forces are assumed to be uncertain. 
In this probabilistic framework, forces are described as a random field or a set of random variables.
Representation of the random objects and propagation of load uncertainties through the model 
are efficiently done through Karhunen–Lo\`{e}ve and polynomial chaos expansions.
We take advantage of using polygonal elements, which have been shown to be effective in suppressing checkerboard patterns and reducing mesh dependency in the solution of topology optimization problems.
Accuracy and applicability of the proposed methodology are demonstrated by means 
of several topology optimization examples. 
The obtained results, which are in excellent 
agreement with reference solutions computed via Monte Carlo method, show that load uncertainties play 
an important role in optimal design of structural systems, so that they must be taken into account to
ensure a reliable optimization process.


\end{abstract}

\section{Introduction}
\label{intro}

Due to new requirements of design associated with the most modern engineering 
applications, mechanical systems with very complex geometrical configurations are 
becoming increasingly common. In this context, some of the most promising 
design approaches are based on topology optimization (TO), which seeks to find the 
best layout for a system, by optimizing the material distribution in a predefined 
design domain \cite{Michell1904p589,Sigmund2004}. The growing popularity
of TO solutions is demonstrated by their wide range of application in various fields 
such as structural mechanics 
\cite{Talischi2009p671,Talischi2012p329,Romero2014p268,
He2014p629,Dapogny2017p933,Luo2017p967},
composite and multi-materials \cite{xia2018p234,Zhang2018}, 
nanotecnology \cite{Nanthakumar2015p97},
fluid mechanics \cite{Pereira2016p1345,Duan2015p40}, 
fluid-structure interaction \cite{Andreasen2013p55}, medicine \cite{Park2018}, etc.

In general, the physical systems underlying TO applications are subjected 
to a series of uncertainties (e.g. unknown loads, geometrical imperfections, 
fluctuations in physical properties, etc) so that, usually, their response is not well predicted by 
the traditional (deterministic) tools of engineering analysis. For this reason, there is 
a consensus among computational engineering experts that uncertainties effects
must be incorporated into any computational predictive model 
\cite{soize2013p2379}. The modeling and quantification of uncertainties is necessary 
in order to predict a possible range of variability for the mathematical model 
response, and to conduct applied tasks, such as analysis and design,
in a robust way \cite{Banichuk2010,soize2017}.  Notice, however, that the majority
of the works in TO area, currently available in the literature, are limited
to deterministic analyses.

The need for robust design and analysis of uncertainties in topological optimization
applications naturally induces the search for computationally efficient frameworks for TO.
For this purpose, TO literature started to take uncertainty quantification (UQ) 
into account over the last decade, as can be seen in several papers addressing 
the two issues \cite{Zhang2017p31,Keshavarzzadeh2017p120,daSilva2017p647,Zhang2017p463,
Putek2016p11,Richardson2016p334,Zhao2014p517,Dunning2013p2656,Jalalpour2013p41,
Tootkaboni2012p263,Asadpoure2011p1031,Chen2010p507,Guest2008p116,Kim2006p1112,WU201636}. 

Some of these works are based on classical techniques for stochastic computation like 
Monte Carlo (MC) method \cite{Chen2010p507,Dunning2013p2656} or series expansion
\cite{daSilva2017p647,Zhao2014p517,Jalalpour2013p41,Asadpoure2011p1031,Guest2008p116}
which, despite of being very simple in conceptual terms, they are limited by the high computational cost, the former,
or very small range of applicability, the latter. These limitations open space for spectral-based approaches
\cite{Kim2006p1112,Tootkaboni2012p263,Richardson2016p334,Zhang2017p31,Putek2016p11,Keshavarzzadeh2017p120},
that use  state-of-the-art tools for representing and propagating uncertainties in computational models,
like Karhunen–Lo\`{e}ve (KL) and generalized polynomial chaos (gPC) expansions.
A recent work by Keshavarzzadeh et. al \cite{Keshavarzzadeh2017p120} presents a non-intrusive gPC strategy to propagate uncertainties in topology optimization problems. They use non-intrusive	polynomial chaos expansion to evaluate low-order statistics of compliance and volume and the uncertainties are considered in the applied loads and also in the geometry of the problems.

The classical formulation for topology optimization, which corresponds to minimize the structural compliance, is commonly carried out on uniform grids consisting of Lagrangian-type finite elements (e.g., linear quads).
However, this choice of discretization, together with density methods, suffer from the well-known numerical instabilities, such as the checkerboard patterns.
Unstructured "Voronoi” meshes, generated from an initial set of random points, have been shown to be effective in suppressing checkerboard patterns \cite{Talischi2009p671}.
Moreover, compared to standard Lagrangian-type uniform grids, polygonal elements are more versatile in discretizing complex domains and in reducing mesh dependency in the solutions of topology optimization \cite{Talischi2009p671,Antonietti2017}.
The geometrical flexibility of the polygonal finite elements also make them very attractive for adaptive mesh refinement schemes in topology optimization problems \cite{Xuan2017,Hoshina2018}.
The computational code used here was developed based on PolyTop \cite{Talischi2012p329}, a MATLAB code for solving topology optimization problems using either structured or unstructured polygonal meshes in arbitrary two-dimensional domains.
The modular structure of \texttt{PolyTop}, where the analysis routine and optimization algorithm are separated from the choice of topology optimization formulation, together with a non-intrusive way of computing the statistical measures, allowed us to implement a robust topology optimization code in a very straightforward way, with only a few modifications in the original \texttt{PolyTop} code.

The aim of this paper is to present a computationally efficient and accurate non-intrusive robust topology optimization approach using polygonal elements.
For this purpose, the \texttt{PolyTop} framework by 
Talischi et al. \cite{Talischi2012p329}, which employs polygonal finite elements in TO, 
is combined with a consistent methodology for stochastic analysis that uses a 
non-intrusive gPC strategy to propagate parameters uncertainties 
through the computational model. This combination generates a framework that 
is computationally efficient for stochastic simulations, and robust to numerical 
instabilities typical of TO problems, such as checkerboards, one-node connections, 
and mesh dependency. The novel approach is used to solve TO problems that seek
to minimize an objective function based on the low-order statistical moments of the 
compliance function of a structure, subjected to uncertainties on the external load,
and satisfying volume constraints.

The remainder of this paper is organized as follows. Stochastic spectral methods are introduced in section 2, 
together with the mathematical formulation and basic steps to obtain KL and gPC expansions. In section 3, the TO 
problem is briefly described, as well as stochastic procedure to propagate uncertainties within TO, 
which is called robust topology optimization (RTO). 
In section 4, numerical examples are presented, and the proposed methodology is compared with 
a reference result obtained with MC method. Finally, some remarks and suggestions for future work 
are presented in section 5.



%

\section{Stochastic spectral methods}
\label{stoch_spec_meth}

\subsection{Preliminary definitions and notation }

Consider a probability space $(\SS, \SF, \PM)$, where
$\SS$ is the sample space, $\SF$ a $\sigma$-field over $\SS$, 
and $\PM: \SF \to [0,1] $ denotes the probability measure.
It is assumed that the distribution $\PD{\randvar{X}} (dx)$ of
any real-valued random variable $\randvar{X}$ in this probability 
space admits a density $x \mapsto \pdf{\randvar{X}} (x)$ with 
respect to $dx$. The set of values where this density
is not zero is dubbed the support of $\randvar{X}$, being
denoted by $\supp{\randvar{X}}$.

In this probabilistic setting, any realization of random variable
$\randvar{X}$ is denoted by $\randvar{X}(\SSpt)$ for $\SSpt \in \SS$,
and the mathematical expectation operator is defined by
\begin{equation}
	\expval{\randvar{X}} = \int_{\R} x \,\, \PD{\randvar{X}} (dx),
	\label{def_expval_op}
\end{equation}
\noindent
so that the mean value and standard deviation of $\randvar{X}$ are given by
$\mean{\randvar{X}} = \expval{\randvar{X}}$ and $\stddev{\randvar{X}} = (\expval{\randvar{X}^2} - \expval{\randvar{X}}^2)^{1/2}$,
respectively. The random variable $\randvar{X}$ is said to be of second-order if
\begin{equation}
	\expval{|\randvar{X}|^2} = \int_{\R} |x|^2 \,\, \PD{\randvar{X}} (dx) \, < \, +\infty.
	\label{def_l2_space}
\end{equation}

The space of all second-order random variables in $(\SS, \SF, \PM)$, denoted by $L_2 (\SS,\PM)$,
is a Hilbert space \cite{brezis2010} equipped with the inner product 
$\innerL{\cdot}{\cdot} : L_2 (\SS,\PM) \times L_2 (\SS,\PM) \to \R$ such that
\begin{equation}
	\innerL{\randvar{X}}{\randvar{Y}} = \expval{\randvar{X} \, \randvar{Y}} =
	\int \, \int_{\R^2} x \, y \,\, \PD{\randvar{X}, \randvar{Y}} (dx,dy),
	\label{def_l2_inner}
\end{equation}
\noindent
where $\PD{\randvar{X}, \randvar{Y}} (dx,dy)$ is the joint distribution
of the random variables $\randvar{X}, \randvar{Y} \in L_2 (\SS,\PM)$.
This inner product induces a norm $\normL{\cdot}: L_2 (\SS,\PM) \to \R$ where
\begin{equation}
	\normL{\randvar{X}} = \left(\innerL{\randvar{X}}{\randvar{X}}\right)^{1/2} = 
	\left( \int_{\R} |x|^2 \,\, \PD{\randvar{X}} (dx) \right)^{1/2}.
	\label{def_l2_norm}
\end{equation}

Further ahead it will also be helpful to consider $L_2 (\mathcal{D})$, 
the set of all real-valued square integrable functions defined on the spatial
domain $\mathcal{D} \subset \R^d, \, d\geq 1$. This set of functions is
also a Hilbert space \cite{brezis2010}, with inner product 
$\innerD{\cdot}{\cdot}: L_2 (\mathcal{D}) \times L_2 (\mathcal{D}) \to \R$ defined by
\begin{equation}
	\innerD{\phi}{\phi'} = 	\int_{\mathcal{D}} \, \phi (\vec{x}) \, \phi' (\vec{x}) \, d\vec{x},
\end{equation}
\noindent
for $\phi, \phi' \in L_2 (\mathcal{D})$.

\subsection{Karhunen–Lo\`{e}ve expansion }
\label{kl_expansion}

The KL expansion \cite{ghanem2003,xiu2010} is one of the most widely 
used and powerful techniques for analysis and synthesis of random fields,
providing a denumerable representation, in terms of the spectral decomposition of the
correlation function, for a random field parametrized by a nondenumerable index
\cite{soize2017,Soize2004p395}.

Let the map $(\vec{x},\SSpt ) \in \mathcal{D} \times \SS \mapsto \randproc{U} \left(\vec{x},\SSpt \right) \in \R$
be an arbitrary real-valued random field, indexed by the spatial coordinate vector 
$\vec{x} \in \mathcal{D} \subset \R^d, \, d\geq 1$, denoted in an abbreviated way as
$\randproc{U} (\vec{x})$ or $\randproc{U}$. By construction, for a fixed $\vec{x} \in \mathcal{D}$, 
$\randproc{U} (\vec{x},\cdot)$ is a real-valued random variable, and 
$\randproc{U} (\cdot,\SSpt)$, for a fixed $\SSpt \in \SS$, is a function 
of $\vec{x}$, dubbed realization of the random field. 

The correlation of $\randproc{U} (\vec{x})$ is the function
$\corr{\randproc{U}}(\cdot,\cdot): \mathcal{D} \times \mathcal{D} \to \R$
defined for any pair of vectors $\vec{x}$ and $\vec{x}'$ by means of
\begin{equation}
	\corr{\randproc{U}} (\vec{x},\vec{x}') = 
	\expval{\randproc{U}(\vec{x}) \, \randproc{U}(\vec{x}')}.
	\label{def_corr_op}
\end{equation}

Suppose that random field $\randproc{U} (\vec{x})$ is second-ordered
and mean-square continuous, properties respectively defined by
\begin{equation}
	\expval{|\randproc{U} (\vec{x})|^2} < \, +\infty, ~~ \forall \, \vec{x} \in \mathcal{D},
\end{equation}
\noindent
and
\begin{equation}
	\lim_{\vec{x}' \to \vec{x}} \, \normL{\randproc{U} (\vec{x}') - \randproc{U} (\vec{x})}^2 = 0.
\end{equation}
\noindent
Under these assumptions, the linear integral operator
\begin{equation}
		\corr{\randproc{U}} \, \phi \, (\vec{x}) = \int_{\mathcal{D}} \corr{\randproc{U}} (\vec{x},\vec{x}') \, \phi (\vec{x}') \, d\vec{x}'
		\label{hs_operator_eq}
\end{equation}
\noindent
defines a Hilbert-Schmidt operator \cite{Soize2004p395,soize2017},
which has denumerable family of eigenpairs 
$\left\lbrace (\lambda_n,\phi_n)\right\rbrace_{n=1}^{+\infty}$ such that
\begin{equation}
		\int_{\mathcal{D}} \corr{\randproc{U}} (\vec{x},\vec{x}') \, \phi_n (\vec{x}') \, d\vec{x}' =
		\lambda_n \, \phi_n (\vec{x}), \qquad \vec{x} \in \mathcal{D},
		\label{fredholm_int_eq}
\end{equation}
where $\lambda_n$ are the eigenvalues and $\phi_n$ the corresponding eigenfunctions 
of the operator defined by Eq.(\ref{hs_operator_eq}). Besides that, the sequence of eigenvalues is such that
$\sum_{n=1}^{+\infty} \lambda_n < +\infty$ and
$\lambda_1 \geq \lambda_2 \geq \cdots \geq \lambda_n \geq \cdots \to 0$;
and the family of functions $\left\lbrace \phi_n \right\rbrace_{n=1}^{+\infty}$ 
defines an orthonormal Hilbertian basis in $L_2 (\mathcal{D})$, i.e.

\begin{equation}
	\innerD{\phi_m}{\phi_n} = \delta_{mn} \,,
\end{equation}
\noindent
where Kronecker delta is such that $\delta_{mn} = 1$ if $m=n$ and
$\delta_{mn} = 0$ for $m \neq n$.

Therefore, applying two standard results of functional analysis \cite{Ciarlet2013},
the theorems of Hilbertian basis and orthogonal projection,
it is possible to show that the random field $\randproc{U} (\vec{x})$ 
admits a decomposition
\begin{equation}
		\randproc{U} (\vec{x}) = \mean{\,\randproc{U}} (\vec{x}) +
		\sum_{n=1}^{+\infty} \sqrt{\lambda_n} \, \phi_n (\vec{x}) \, \xi_n \,,
		\label{kl_decomp}
\end{equation}
\noindent
where $\left\lbrace \xi_n \right\rbrace_{n=1}^{+\infty}$ is a family of
random variables defined by
\begin{equation}
		\xi_n = \frac{1}{\sqrt{\lambda_n}}  \, \innerD{ \randproc{U} - \mean{\,\randproc{U}} } {\phi_n},
\end{equation}
which are centered (zero mean) and mutually uncorrelated i.e.
\begin{equation}
		\mean{\xi_n} = 0,
		\qquad \mbox{and} \qquad
		\expval{\xi_m \, \xi_n} = \delta_{mn} \,.
		\label{kl_uncorrelated}
\end{equation}

A finite dimensional approximation for $\randproc{U} (\vec{x})$, 
denoted by $\randproc{U}^{\nu_{kl}} (\vec{x})$, is constructed by truncation of 
the series in Eq.(\ref{kl_decomp}) i.e.
\begin{equation}
		\randproc{U}^{\nu_{kl}} (\vec{x}) = \mean{\,\randproc{U}} (\vec{x}) +
		\sum_{n=1}^{\nu_{kl}} \sqrt{\lambda_n} \, \phi_n (\vec{x}) \, \xi_n,
		\label{kl_decomp_finite}
\end{equation}
\noindent
where the integer $\nu_{kl}$ is chosen such that 
\begin{equation}
		\texttt{energy}(\nu_{kl}) = \frac{\sum_{n=1}^{\nu_{kl}} \lambda_n}{\sum_{n=1}^{+ \infty} \lambda_n} \geq \tau,
		\label{kl_conv_crit}
\end{equation}
\noindent
for a heuristically chosen threshold $\tau$ (e.g. $\tau=90\%$). In practice, 
as a closed formula for $\lambda_n$ is not available in general, $\texttt{energy}(\nu_{kl})$ 
is estimated using a finite (but large) number of eigenvalues, instead of an infinite 
quantity. This procedure is justified in light of the eigenvalues decreasing property.

One of the main difficulties to apply KL expansion to discrete random fields is the determination 
of the eigenvalues and corresponding eigenfunctions of the correlation function. Analytical 
solutions for the Fredholm integral equation in (\ref{fredholm_int_eq}) are almost never available. 
However, for some special cases, such as exponential and Gaussian autocovariance functions, 
an analytical solution can be obtained by converting the integral equation into a differential 
equation through successive derivatives \cite{ghanem2003,xiu2010}.

Several numerical methods can be used to solve the eigenvalue problem of Eq.(\ref{fredholm_int_eq}),
such as the direct method, projection methods,
among others \cite{Atkinson2009,Betz2014p109}. In this study, the direct method is employed
to transform the Fredholm integral equation into a finite dimensional eigenvalue problem, whose the solution 
provides an approximation for the desired eigenvalues/eigenvectors of the infinite dimensional problem.

In this numerical procedure, a set of $M$ realizations of the random field $\randproc{U}(\vec{x})$ and 
its mean function $\mu_{\randproc{U}}(\vec{x})$ 
are numerically generated\footnote{These numerical realizations are defined in a computational mesh $\vec{x}_1,\vec{x}_2,\cdots,\vec{x}_n$.}
and grouped into the matrices

\begin{equation}
\begin{array}{rcl}
\randproc{\bm U} &=&
\begin{bmatrix}
U^1(\vec{x}_1) & U^2(\vec{x}_1) &  \dots  & U^M(\vec{x}_1) \\
U^1(\vec{x}_2) & U^2(\vec{x}_2) &  \dots  & U^M(\vec{x}_2) \\
\vdots & \vdots &\ddots & \vdots \\
U^1(\vec{x}_n) & U^2(\vec{x}_n) & \dots  & U^M(\vec{x}_n)
\end{bmatrix},
\end{array}
\begin{array}{rcl}
\bm\mu &=&
\begin{bmatrix}
\mu_{\randproc{U}}(\vec{x}_1) & \mu_{\randproc{U}}(\vec{x}_1) &  \dots  & \mu_{\randproc{U}}(\vec{x}_1) \\
\mu_{\randproc{U}}(\vec{x}_2) & \mu_{\randproc{U}}(\vec{x}_2) &  \dots  & \mu_{\randproc{U}}(\vec{x}_2) \\
\vdots & \vdots &\ddots & \vdots \\
\mu_{\randproc{U}}(\vec{x}_n) & \mu_{\randproc{U}}(\vec{x}_n) & \dots  & \mu_{\randproc{U}}(\vec{x}_n)
\end{bmatrix} 
\end{array},
\end{equation}
which are used to define the zero mean matrix 
$\boldsymbol{\hat{U}}=\randproc{\bm U}-\bm\mu$. Then, the correlation 
matrix is estimated with the aid of

\begin{equation}
\corr{\randproc{\bm{\hat U}}} = \frac{1}{M}\bm{\hat{U}}\bm{\hat{U}}^T,
\end{equation}
and the discrete eigenvalue problem
\begin{equation}
\corr{\randproc{\bm{\hat U}}}  \boldsymbol{\Phi} = \boldsymbol{\Lambda} \boldsymbol{\Phi},
\end{equation}
\noindent
is solved to obtain the matrices $\boldsymbol{\Phi}$ and $\boldsymbol{\Lambda}$, 
which present approximations for the first $M$ eigenfunctions/eigenvalues on the 
columns/main diagonal.

For further details on theoretical and practical aspects of KL expansion
the reader is encouraged to see \cite{lemaitre2010,bellizzi2012p3509,Perrin2013p607,iaccarino2015}.

\subsection{Generalized polynomial chaos expansion }

The gPC expansion is a theoretical tool used to construct representations for random fields, with a
denumerable or nondenumerable set of index, in terms of a denumerable collection of 
random variables weigthed by deterministic coefficients \cite{soize2017,Soize2004p395}. 
It was introduced in the engineering 
community by R. Ghanem \cite{ghanem2003,ghanem1989p1035,Ghanem1999p19} 
as a tool to compute approximate responses for problems involving random fields with 
unknown distribution and since the early 2000s, especially after the work of Xiu and 
Karniadakis \cite{xiu2002p619}, it has been used in many applications of computational 
stochastic mechanics, a trend that should increase \cite{Stefanou2009p1031}.

For the sake of theoretical development,
consider a second-ordered random variable $\randvar{U}: \SS \to \R$
which can be written in terms of a (possibly infinity) set of independent random variables 
$\xi_1(\SSpt), \xi_2(\SSpt), \xi_3(\SSpt), \cdots$. Collecting these independent 
variables into the random vector $\bm{\xi}(\SSpt) = (\xi_1(\SSpt), \xi_2(\SSpt), \xi_3(\SSpt), \cdots)$,
dubbed the germ, it is possible to rewrite the original variable in the parametric form
$\randvar{U} = \randvar{U}(\bm{\xi})$.

In this context of second-ordered random variables, gPC expansion theory says 
that such random variable $\randvar{U}$ admits a spectral representation
\begin{equation}
	\label{gpc_eq}
	\randvar{U}(\bm{\xi}) = \sum_{n=0}^{+\infty} u_n \, \psi_n (\bm{\xi}),
\end{equation}
where the set of orthonormal polynomials $\{\psi_0,\psi_1,\psi_2,\cdots \}$
is a basis for $L_2 (\SS,\PM)$, and the deterministic coefficients 
$u_0, u_1, u_2, \cdots$ correspond to the coordinates of $\randvar{U}$ 
in this infinite dimensional base. By definition $\psi_0(\bm{\xi}) = 1$, 
and due to the orthogonality property $\innerL{\psi_m}{\psi_n} = \delta_{mn}$,
one has
\begin{eqnarray}
	\mean{\randvar{U}} & = & \expval{\randvar{U}}\\ \nonumber
								   & = & \expval{\psi_0 \, \randvar{U}}\\ \nonumber
								   & = & \sum_{n=0}^{+\infty} u_n \, \innerL{\psi_0}{\psi_n}\\ \nonumber
								   & = & u_0,
\end{eqnarray}
i.e., the mean of $\randvar{U}$ is equal to the coordinate $u_0$ of gPC expansion.
Besides that, it is easy to see that $\expval{\psi_n} = 0, \,\, n \geq 1$.
Thus, it can also be shown that standard deviation of $\randvar{U}$ can be written as
\begin{eqnarray}
%
\sigma_{U}^2 & = & E \left\lbrace U^2 \right\rbrace  - E \left\lbrace U \right\rbrace^2 \\ \nonumber
& = & E \left\lbrace \left(\sum_{n=1}^{+\infty} u_n \, \psi_n(\bm{\xi}) \right)^2 \right\rbrace \\ \nonumber
& = & \sum_{n=1}^{+\infty} u_n^2.
\end{eqnarray} 

The gPC expansion has attracted the attention of many researchers due to its rapid 
convergence property and its capability to estimate statistical moments, which allows 
this method to efficiently reduce computational effort in highly nonlinear engineering 
design applications \cite{lemaitre2010,iaccarino2015}. 

Of course, for any purpose of numerical computation, it is necessary to
parametrize the random variable $\randvar{U}$ with a finite number $\nu_{rv}$ 
of independent random variables, i.e., $\bm{\xi} = (\xi_1, \xi_2, \cdots, \xi_{\nu_{rv}})$,
and restrict up to $p_{pc}$ the order of the polynomials in the basis, so that
the series in Eq.(\ref{gpc_eq}) is truncated with $\nu_{pc}+1 = (\nu_{rv}+p_{pc})! /(\nu_{rv}! \, p_{pc}!)$ terms, 
giving rise to the approximation
\begin{equation}
	\randvar{U}^{\nu_{pc}}(\bm{\xi}) = \sum_{n=0}^{\nu_{pc}} u_n \, \psi_n (\bm{\xi}),
	\label{gpc_trunc_eq}
\end{equation}
which is mean-square convergent to $\randvar{U}$ when $p_{pc}, \nu_{rv} \to +\infty$.

In practice, $\nu_{rv}$ is obtained from the stochastic modeling and $p_{pc}$ is specified such that a compromise between accuracy and efficiency can be established.
In this work, the gPC polynomial order is selected such that the low-order statistics estimations become invariant for increasing values of $p_{pc}$.

This spectral expansion can be easily extended to a second-ordered 
random field $\randproc{U}: \mathcal{D} \times \SS \to \R$, considering this field 
as a family of random variables $\randproc{U}(\vec{x},\cdot) \in L_2(\SS,\PM)$,
parameterized by the index $\vec{x}$, and  by letting the deterministic coefficients of the 
gPC expansion depend on $\vec{x}$, i.e.
\begin{equation}
	\randvar{U}(\vec{x},\bm{\xi}) = \sum_{n=0}^{+\infty} u_n (\vec{x}) \, \psi_n (\bm{\xi}),
	\label{gpc_randfield_eq}
\end{equation}
where the coordinates are now a deterministic function of $\vec{x}$.
As in the case of a random variable, the truncation of the series in Eq.(\ref{gpc_randfield_eq})
results in the approximation
\begin{equation}
	\randvar{U}^{\nu_{pc}}(\vec{x},\bm{\xi}) = \sum_{n=0}^{\nu_{pc}} u_n (\vec{x}) \, \psi_n (\bm{\xi}).
\end{equation}

Basically, two approaches for implementing the gPC expansion are available in the literature,
one called the intrusive, and another one dubbed non-intrusive. 
The intrusive approach is based on a stochastic version of Galerkin method \cite{xiu2010}, 
and consists in modifying the deterministic model in order to take into account the uncertainty 
propagation. The main disadvantage of this formalism is related to the level of difficulty in modifying 
the code associated with the deterministic model, particularly if commercial software is being used, 
due to code access restrictions \cite{soize2017,Ghanem2017chap}. 

On the other hand, the non-intrusive approach requires no modification in the deterministic 
model and therefore, can be treated as a black box \cite{soize2017,Ghanem2017chap}. 
In this case, a probabilistic collocation 
approach, based on a sparse grid method \cite{Eldred2009}, can be used to estimate the 
coefficients (coordinates) of the expansion. In this work, a different non-intrusive approach based on linear regression is employed, where the random field (variable) 
of interest is evaluated in a finite set of $\nu_{gq}$ possible realizations of the 
germ $\bm{\xi}$, and thus the coefficients of the expansion are obtained through 
$\vec{u} = (\bm{\Psi}^{\,T} \, \bm{\Psi})^{-1}\bm{\Psi}^{\,T} \, \vec{U}$,
the solution of the mean-square problem
\begin{equation}
	\bm{\Psi} \, \vec{u} \,\, \approx \,\, \vec{U},
	\label{gpc_regression}
\end{equation}
where
\begin{equation}
\underbrace{
\left[
\begin{array}{ccc} 
\psi_0 \left(\bm{\xi}_1 \right) & \cdots & \psi_{\nu_{pc}} \left( \bm{\xi}_1 \right)\\
\vdots & \ddots & \vdots\\
\psi_0 \left( \bm{\xi}_{\nu_{gq}} \right) &\cdots & \psi_{\nu_{pc}} \left( \bm{\xi}_{\nu_{gq}} \right)\\
\end{array}
\right]
}_{\bm{\Psi}}
\,\,
\underbrace{
\left[\begin{array}{c} u_0\\ u_1\\ \vdots\\ u_{\nu_{pc}}\\\end{array}\right] 
}_{\bm{u}}
\approx
\underbrace{
\left[\begin{array}{c} U(\vec{x},\bm{\xi}_1)\\ \vdots\\ U(\vec{x},\bm{\xi}_{\nu_{gq}})\\
\end{array}
\right].
}_{\bm{U}}
\end{equation}

For further information about the basic aspects of gPC expansion the reader is encouraged 
to see the references \cite{ghanem2003,xiu2010,lemaitre2010,iaccarino2015,Ghanem2017chap,kundu2014stochastic}, 
and for more advanced topics \cite{Soize2004p395,Soize2010p2820,Soize2015p34,soize2017p201,kundu2018probabilistic}.


\section{Topology optimization}
\label{rob_top_opt}

\subsection{Classical topology optimization }

The main objective of TO is to find the optimal distribution of materials, for every point 
$\vec{x}$ in a given design domain $\Omega \subset \R^d$, $d=2$ or $3$, which maximize a 
certain performance measure subjected to a set of design constraints, i.e.,
to determine which regions in $\Omega$ should not present material (void regions), 
and obtain the final topology of the structure. 

By convention, points where material exists are 
represented by a density value of 1, otherwise, the density value is 0. Note that, in this way,
one has an integer-programming problem, where the distribution of material 
is defined by the density map 
$\vec{x} \in \Omega \mapsto \rho(\vec{x}) \in \{0,1\}$, for
\begin{equation}
	\rho(\vec{x}) = 
	\begin{cases}
       1 & \mbox{if} ~~ \vec{x} ~~ \mbox{is structural member},\\
       0 & \mbox{if} ~~ \vec{x} ~~ \mbox{is void}.
     \end{cases}
\end{equation}

In this paper the optimization problem seeks to minimize an objective function
defined by the continuum structure compliance, denoted here by $c$,
subject to a constraint on the final volume of the structure
and satisfying the equilibrium equations for a linear Hookean solid
material.
This formulation, which is equivalent to maximize the structural stiffness subject to  
the same constraints
(see reference \cite{Sigmund2004} for further details),
 can be stated as
\begin{equation}
\begin{aligned}
\min_{\rho} & & c \left(\vec{u}(\rho),\rho \right) & = & \int_{\Omega} \frac{1}{2} \, \bm{\sigma} : \bm{\epsilon} \,\, d\Omega,~~~~\\
s.t.              & &                                           v(\rho) & = &  \int_{\Omega} \rho(\vec{x}) \, d\Omega \leq v_{S},
\end{aligned}
\end{equation}
where $\bm{\sigma}$ and $\bm{\epsilon}$ represent the tensors of stress and strain, respectively,
$v_{S}$ is a specified upper bound on the optimized structure volume,
and the map $\vec{x} \in \Omega \mapsto \vec{u}(\rho(\vec{x})) \in \R^{d}$ is the continum structure 
displacement, parametrized by $\rho$ and implicitly defined by the elasticity equations
\begin{equation}
\begin{array}{rcl}
\nabla \cdot \bm{\sigma}(\vec{u})   & = &  \bm{0} ,     \\
\bm{\sigma}(\vec{u})                         & = & \bm{\sigma}(\vec{u})^T,             \\ 
\bm{\epsilon}(\vec{u})                       & = &  \frac{1}{2} \left( \nabla \vec{u} + \nabla \vec{u}^T \right),    \\ 
\bm{\sigma}(\vec{u})                         & = &  \bm{\mathcal{C}}(\rho) : \bm{\epsilon}(\vec{u}),      
\end{array}
\end{equation}
\noindent
and the boundary conditions
\begin{equation}
\begin{array}{rcl}
\bm{\sigma}(\vec{u}) \cdot \vec{n}  & = &  \bm{t}      \,~ \mbox{in}\, \Gamma_N,     \\ 
\vec{u}                             & = &  \bm{0}      \,~ \mbox{in}\, \Gamma_D,    
\end{array}
\end{equation}
\noindent
where $\Gamma_D$ is the partition of $\partial \Omega$ on which the displacements are prescribed, $\Gamma_N$ is the complimentary
partition  of $\partial \Omega$ on which tractions $\vec{t}$ are prescribed such that $\overline{\Gamma_D} \cup \overline{\Gamma_N} = \partial \Omega$ and ${\Gamma_D} \cap {\Gamma_N} = \emptyset$ and $\bm{\mathcal{C}}(\rho)$ is the $4th$ order stiffness tensor that depends on the density function $\rho$.
As posed, finding $\rho$ and $\bm{\mathcal{C}}(\rho)$ becomes a large integer programming problem, which can be impractical to solve. Thus, we recast $\rho$ as a continuous scalar field, $\rho(\vec{x}) \in [0,1]$.
In order to recover the binary nature of the problem, the SIMP~\cite{Sigmund2004} model is employed and the stiffness tensor can be expressed as

\begin{equation}
\bm{\mathcal{C}}(\rho)=\left[\varepsilon+(1-\varepsilon) \rho^p \right] \bm{\mathcal{C}}^0,
\end{equation}

\noindent
where $p>1$ is the penalty parameter, $\bm{\mathcal{C}}^0$ is the elasticity tensor of the constituent material,
and $0<\varepsilon\ll1$ is a positive parameter ensuring well-posedness of the governing equations.

In terms of computational implementation, the domain is splited into
$N$ elementary regions, i.e., $\Omega = \cup_{e=1}^{N} \Omega^e$, and
the finite element method is employed
for the solution of 
the elasticity equations. Thus, the following finite dimensional version of the optimization problem is considered
\begin{eqnarray}
\label{nand_form_eq}
\min_{\bm{\rho}} & & C(\bm{\rho}) = \vec{F}^{\,\,T} \, \vec{U} (\bm{\rho}),\\
\label{eqn_vol_const} \mbox{s.t.} & & V(\bm{\rho}) = \sum_{e=1}^{N} \rho_e \, |\Omega^e| - v_{S} \leq 0, \\
\mbox{with} & & \mat{K} (\bm{\rho}) \, \vec{U}(\bm{\rho}) = \vec{F},
\end{eqnarray}
where $\bm{\rho}=(\rho_1,\rho_2,\cdots,\rho_N)$
is a discretized version of the density map, $|\Omega^e|$ denotes the volume
of the element $e$, $\vec{U}$ is the discrete displacement vector, 
parametrized by $\bm{\rho}$ and implicitly defined by the equilibrium equation,
$\vec{F}$ is the global loading vector and $\mat{K}$ represents the global stiffness matrix, 
which is also dependent on $\bm{\rho}$.


In order to use gradient-based optimization algorithms to solve the above formulation,
gradients of the objective function as well as the volume constraint function
are needed.
The sensitivity of the objective function $C(\bm{\rho})$ with respect to the design variables $\bm{\rho}$, is expressed in component form as

\begin{equation}
\label{eqn_comp_sens}
\frac{\partial C}{\partial \rho_e} = - \vec{U}^T \frac{\partial \mat{K}}{\partial \rho_e} \vec{U}.
\end{equation}


The gradient of the volume constraint function $V(\bm{\rho})$ with respect
to the design variable $\rho_e$ is given as

\begin{equation}
\label{eqn_vol_sens}
\frac{\partial V}{\partial \rho_e} = |\Omega^e|.
\end{equation}

During the solution of a TO problem it is very common to deal with numerical 
anomalies, such as checkerboards, which are traditionally treated through the 
use of higher-order elements or filters \cite{Sigmund1998p68,Bruggi2008p1819}.
However, Talischi et al. \cite{Talischi2012p329} have shown that the use of the
\texttt{PolyTop} framework, which employs polygonal finite elements,
can naturally address the checkerboard problem. Besides that, this approach
also allows flexibility in the optimization strategy to be used, being compatible 
with the classical approaches based on the optimality criteria (OC) \cite{Kikuchi1988p197}
and the method of moving asymptotes (MMA) \cite{Svanberg1987p359}.

\begin{figure*}[h!]
\centering
\includegraphics[scale=0.8]{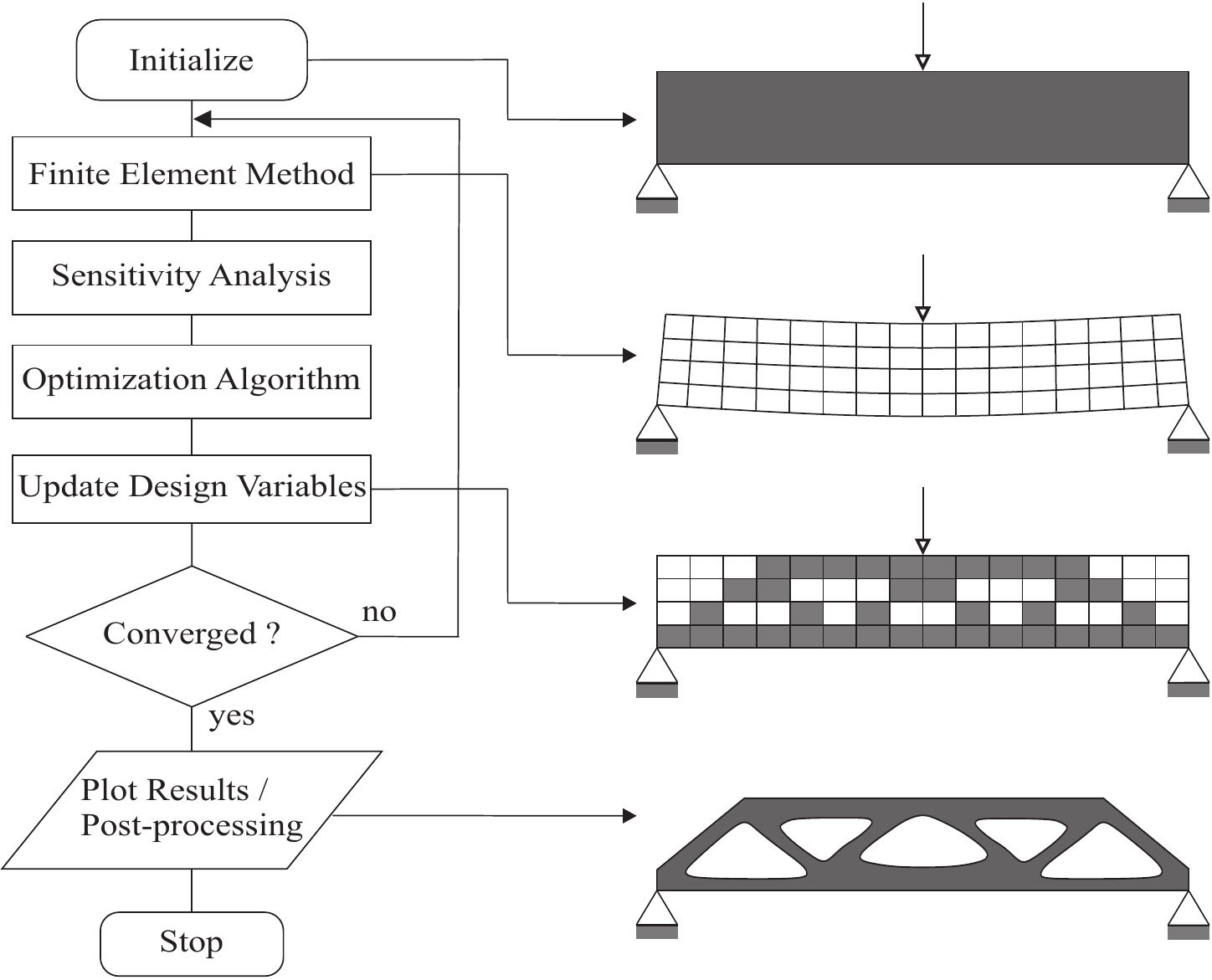}
\caption{Overview of the classical topology optimization procedure (adapted from \cite{Sigmund2004}).}
\label{to_algorithm_fig}
\end{figure*}

An overview of the classical TO procedure, used in \texttt{PolyTop} framework
to obtain an optimal design, is illustrated in Figure~\ref{to_algorithm_fig}. 
The sensitivity analysis step described in this schematic is explained in details 
in section~\ref{sens_analysis}.

\subsection{Robust design optimization }
\label{rdo_section}

\begin{figure*}
\centering
\includegraphics[scale=0.6]{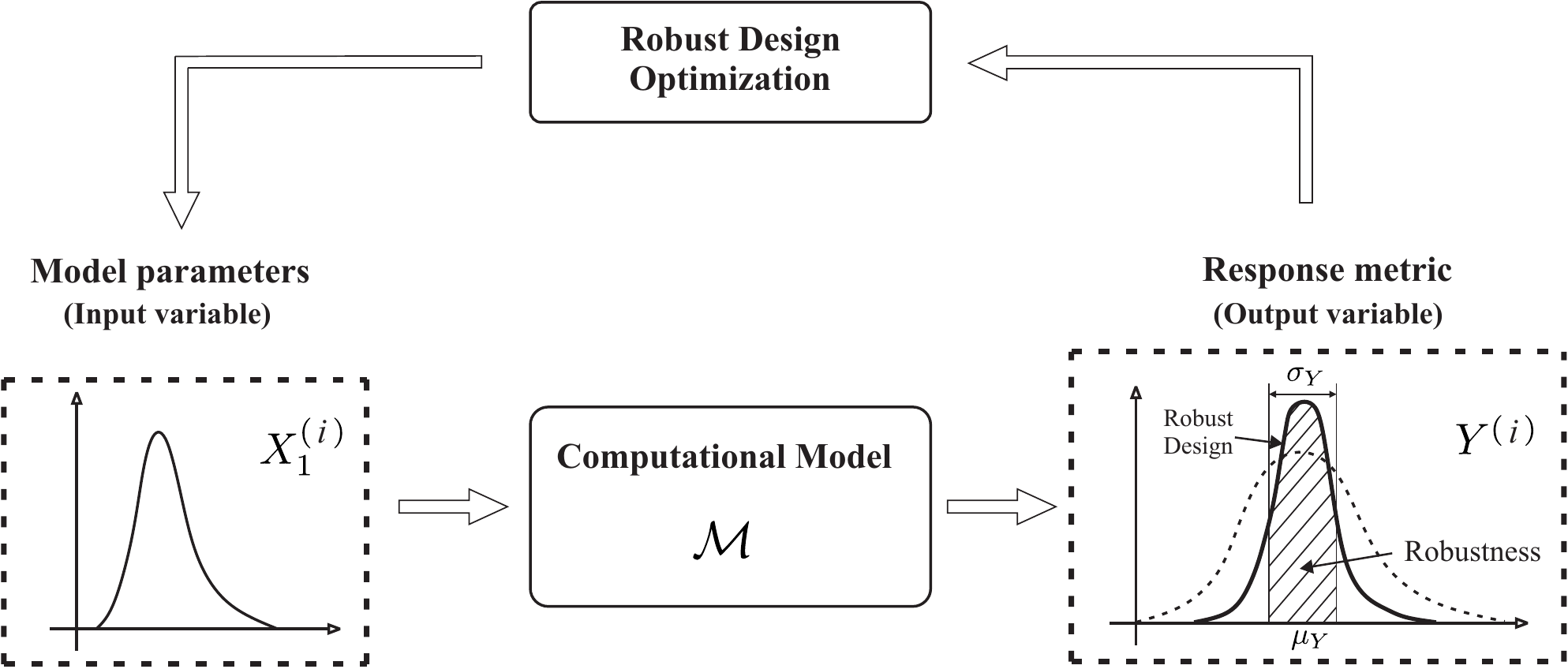}
\caption{General overview of the robust design optimization procedure.}
\label{rto_fig}
\end{figure*}

Robust optimization, also known as robust design optimization (RDO), is a mathematical 
procedure that simultaneously addresses optimization and robustness analysis, obtaining an optimal 
design that is less susceptible to variabilities (uncertainties) in the system parameters. In contrast to 
conventional optimization, that is deterministic, RDO considers parameters that are random
so that it consists of a stochastic problem. The general overview of RDO is explained in Figure~\ref{rto_fig},
which shows a computational model where the input is subjected to uncertainties
--- that can be in material or geometrical properties, loadings, boundary conditions etc. --- and, 
therefore, the model response has a certain probability distribution. This distribution is used to 
compute some kind of statistical response of the system, which is conveniently used to update 
the model input, in order to reduce the output uncertainty.

In this context, it is essential to understand the mathematical definition of robustness, i.e., the choice 
of the robustness measure that is generally expressed by the combination of statistical properties of the 
objective function. Several definitions of measures of robustness have been proposed in literature 
\cite{beyer2007p3190,Birge2011,Doltsinis2004p2221,Shin2011p248} and the weighted sum of both the mean 
and the standard deviation of the objective function is often considered. The tradeoff between these two 
statistical measures gives rise to a final design that is less sensitive to parameters variations,
i.e., a kind of robust design.

\subsection{Robust topology optimization }

In order to increase the optimal design robustness, the concept of robust optimization 
described in section~\ref{rdo_section} can be applied to TO. This possibility is addressed
in this paper where variabilities in the external loading acting on the structure of interest
are taken into account. Thus, the force vector and the compliance function become random
objects, more precisely, a random vector $\vec{F}(\SSpt)$ and a random variable 
$C(\bm{\rho},\SSpt)$, both defined on the probability space $(\SS,\SF,\PM)$. 

For the sake of computational implementation,
these objects are parametrized by a set of $\nu_{rv}$ suitable random variables that are lumped 
into the germ $\bm{\xi}(\SSpt) = \left(\xi_1(\SSpt), \xi_2(\SSpt), \cdots, \xi_{\nu_{rv}}(\SSpt) \right)$
so that force vector and compliance can be expressed as $\vec{F}(\bm{\xi})$ and $C(\bm{\rho},\bm{\xi})$.

A straightforward measure of structural performance (robustness of the objective function) in RDO framework
is given by the mean of the compliance $\mean{C(\bm{\rho},\bm{\xi})}$. However, the final design may still be 
sensitive to the fluctuation due to external loading uncertainties and this may give rise to the need for a more 
robust design \cite{Dunning2013p2656}. Then, the standard deviation of the compliance $\stddev{C(\bm{\rho},\bm{\xi})}$
is also introduced into the formulation of the structural performance measure $\tilde{C}$,
so that it is a linear combination between the mean and standard deviation of the random compliance $C(\bm{\rho},\bm{\xi})$.
By combining these two statistics it is possible to improve the design
by minimizing the variability of the structural 
performance, satisfying the volume constraint. 

This procedure, called robust topology optimization (RTO), can be mathematically formulated as
\begin{eqnarray}
\label{robust_to_eq}
\min_{\bm{\rho}} & & \tilde{C}(\bm{\rho}) = \mean{C(\bm{\rho},\bm{\xi})} + w \, \stddev{C(\bm{\rho},\bm{\xi})},\\ \nonumber
s.t. & & V(\bm{\rho}) = \sum_{e=1}^{N} \rho_e \, |\Omega^e| \leq v_{S},
\end{eqnarray}
which depends on the weight $w\geq0$ and on the random displacement map $\vec{U}(\bm{\rho},\bm{\xi})$,
implicitly defined by the random equilibrium equation 
\begin{equation}
\label{eqn_fem}
\mat{K} (\bm{\rho}) \, \vec{U}(\bm{\rho},\bm{\xi}) = \vec{F}(\bm{\xi}).
\end{equation} 

The RTO problem defined in (\ref{robust_to_eq}) can be solved by considering non-intrusive methods for stochastic 
computation. The basic idea of non-intrusive methods is to use a set of deterministic model evaluations to construct 
an approximation of the desired (random) output response. The deterministic evaluations are obtained for a finite
set of realizations of parameter $\bm{\xi}$ with the aid of a deterministic solver (e.g. finite element code), that is 
used as a black box. Thus, non-intrusive methods offer a very simple way to propagate uncertainties in complex 
models, such as structural optimization, where only deterministic solvers are available. In this study the focus is 
on two non-intrusive techniques, namely, MC simulation \cite{kroese2011,rubinstein2016} and 
gPC expansion \cite{xiu2010,Ghanem2017chap}.

\subsection{Low-order statistics for compliance }

Monte Carlo (MC) method is one of the simplest crude techniques for stochastic simulation and may be
used to construct mean-square consistent and unbiased estimations (approximations)
--- see \cite{kroese2011} for details --- for $\mean{C} = \mean{C(\bm{\rho},\bm{\xi})}$ 
and $\stddev{C} = \stddev{C(\bm{\rho},\bm{\xi})}$, respectively defined by
\begin{equation}
	\estim{\mu}_{C} =
	\frac{1}{\nu_{mc}} \ \sum_{n=1}^{\nu_{mc}} C_n, 
\end{equation}
and
\begin{equation}
	\estim{\sigma}_{C} =  
	\left(\frac{1}{\nu_{mc}-1} \ \sum_{n=1}^{\nu_{mc}} \left( C_n - \estim{\mu}_{C} \right)^2\right)^{1/2},
\end{equation}
where $C_n = C(\bm{\rho},\bm{\xi}_n)$,
$\bm{\xi}_n$ is the n-th realization of the germ $\bm{\xi}$ and
$\nu_{mc}$ denotes the number of MC realizations.

Despite its simplicity, the slow convergence rate of MC method ($\sim 1/\sqrt{\nu_{mc}}$) 
usually makes it a very expensive stochastic solver in terms of computational cost,  
particularly for TO problems, where a large number of deterministic model resolutions needs to be 
obtained to achieve an adequate response characterization. For this reason, a gPC procedure 
for low-order statistics estimation is also considered in this work.

Using the gPC approach, an spectral representation of the compliance function can be written as
\begin{equation}
	C(\bm{\rho},\bm{\xi}) = \sum_{n=0}^{+\infty} c_n(\bm{\rho}) \, \psi_n (\bm{\xi}),
\end{equation}
in a way that, because of properties $\expval{\psi_0} = 1$ and $\expval{\psi_n} = 0, \,\, n \geq 1$, 
the mean value of $C(\bm{\rho},\bm{\xi})$ writes as
\begin{equation}
	\mean{C} = \expval{C(\bm{\rho},\bm{\xi})} = c_0 (\bm{\rho}),
\end{equation}
where an approximation for the PCE coefficient $c_0$ is obtained from the linear regression (\ref{gpc_regression}).
This procedure induces a Gaussian quadrature estimation of $\mean{C}$,
defined by the estimator
\begin{equation}
	\estim{\mu}_{C}^{\,\,'} = \sum_{j=1}^{\nu_{gq}} W_j \, C_j,
	\label{gpc_mean_estim}
\end{equation}
where $ C_j =  C(\bm{\rho},\bm{\xi}_j)$ corresponds to the evaluation of the
compliance at the Gauss points and the quadrature weights $W_j = \Psi_{\,1j}^{\dagger}$ are given by 
the first line entries of $\mat{\Psi}^{\dagger} = (\mat{\Psi}^T \, \mat{\Psi})^{-1} \, \mat{\Psi}^T$, 
the pseudoinverse  of the $\nu_{gq} \times \nu_{pc}$ regression matrix $\mat{\Psi}$.

By definition, the standard deviation of compliance is written as
\begin{equation}
	\stddev{C} = \left(\expval{C(\bm{\rho},\bm{\xi})^2} - \expval{C(\bm{\rho},\bm{\xi})}^2\right)^{1/2},
\end{equation}
so that a procedure similar to that used to construct the estimator of Eq.(\ref{gpc_mean_estim}) 
can be adopted now to propose
\begin{equation}
	\estim{\sigma}_{C}^{\,\,'} = \left(\sum_{j=1}^{\nu_{gq}} W_j \, C_j^{2} - \left(\estim{\mu}_{C}^{\,\,'} \right)^2\right)^{1/2},
	\label{gpc_stddev_estim}
\end{equation}
as an estimator for $\stddev{C}$.

The gPC-based estimators defined by Eqs.(\ref{gpc_mean_estim}) and (\ref{gpc_stddev_estim})
provide a very accurate and efficient framework for estimation of the compliance low-order statistics,
that demands a small number of deterministic model evaluations, once in practice
$\nu_{gq} \ll \nu_{mc}$.

\subsection{Sensitivity analysis }
\label{sens_analysis}

The partial derivative of the objective function $\tilde{C}$, defined in the
optimization problem (\ref{robust_to_eq}), with respect to the element density function $\rho_e$
is given by
\begin{equation}
	\frac{\partial \tilde{C}}{\partial \rho_e} = 
	\frac{\partial \mean{C}}{\partial \rho_e}  + 
	w\, \frac{\partial \stddev{C}}{\partial \rho_e},
	\label{gpc_mean_sens}
\end{equation}
where the partial derivatives on the right hand side can be approximated, via crude MC,
with the aid of the estimators
\begin{equation}
    \frac{\partial \mean{C}}{\partial \rho_e} \approx
	\estim{\frac{\partial \mean{C}}{\partial \rho_e}} =
	\frac{1}{\nu_{mc}} \ \sum_{n=1}^{\nu_{mc}} \frac{\partial C_n}{\partial \rho_e}
\end{equation}
and
\begin{equation}
\frac{\partial \sigma_{C}}{\partial \rho_e} \approx
\widehat{\frac{\partial \sigma_{C}}{\partial \rho_e}} =
\frac{1}{\left( \nu_{mc}-1\right)\widehat{\sigma}_{C}} \
\left( \left(\sum_{n=1}^{\nu_{mc}}  C_{n} \, \frac{\partial C_{n}}{\partial \rho_e}\right) - 
\nu_{mc} \, \widehat{\mu}_{C} \, \widehat{\frac{\partial \mu_{C}}{\partial \rho_e}}  \right).
\end{equation}

However, from the point of view of computational cost, it is more efficient to obtain these
sensitivity coefficients using the gPC estimators, i.e.
\begin{equation}
   \frac{\partial \mean{C}}{\partial \rho_e} \approx
	\estim{\frac{\partial \mean{C}}{\partial \rho_e}}^{\,\,'} = \sum_{j=1}^{\nu_{gq}} W_j \, \frac{\partial C_j}{\partial \rho_e}
\end{equation}
and
\begin{equation}
    \frac{\partial \stddev{C}}{\partial \rho_e} \approx
	\estim{\frac{\partial \stddev{C}}{\partial \rho_e}}^{\,\,'} = \frac{1}{\estim{\sigma}_{C}^{\,\,'}} \,
	\left( \left(\sum_{j=1}^{\nu_{gq}} W_j \, C_j \, \frac{\partial C_j}{\partial \rho_e} \right) - \estim{\mu}_{C}^{\,\,'} \, \estim{\frac{\partial \mean{C}}{\partial \rho_e}}^{\,\,'} \right),
	\label{gpc_stddev_sens}
\end{equation}
obtained from Eqs.(\ref{gpc_mean_estim}) and (\ref{gpc_stddev_estim})
by differentiation with respect to $\rho_e$.

\subsection{Algorithm for robust topology optimization }

The results obtained from the TO algorithm, i.e., the compliance and sensitivities, 
are used to compute statistical measures in a non-intrusive way. Therefore, the RTO algorithm, 
for problems with uncertain loading, can be described as follows:

\begin{enumerate}
	\item Topology Optimization Data: define finite element model, set optimizer and underlying numerical and control parameters;
	\item Stochastic Model: parametrize aleatory objects with a set of independent random variables defined by the germ $\bm{\xi}$. Choose an appropriate family of orthogonal polynomials, define weight factor $w$ and gPC~order~$p_{pc}$;
	\item Objective Function
	\begin{itemize}
        \item for each germ realization $\bm{\xi}_n$ perform finite element analysis using Eq.(\ref{eqn_fem}) and	compute compliance sensitivities with Eq.(\ref{eqn_comp_sens});
	    \item Compute gPC coefficients from Eq.(\ref{gpc_regression});
	    \item Compute statistical estimates for mean and standard deviation with aid
	    of Eqs.(\ref{gpc_mean_estim}) and (\ref{gpc_stddev_estim});
	    \item Compute the sensitivity of the objective function  from Eq.(\ref{gpc_mean_sens}).
    \end{itemize}
	\item Constraint Function: compute volume constraint using Eq.(\ref{eqn_vol_const}) and its sensitivity with Eq.(\ref{eqn_vol_sens});
	\item Update the design variables $\bm{\rho}$  according to the optimizer. Repeat from step 3 until 
	convergence is achieved;
\end{enumerate}

\begin{figure*}
\centering
\includegraphics[scale=1.0]{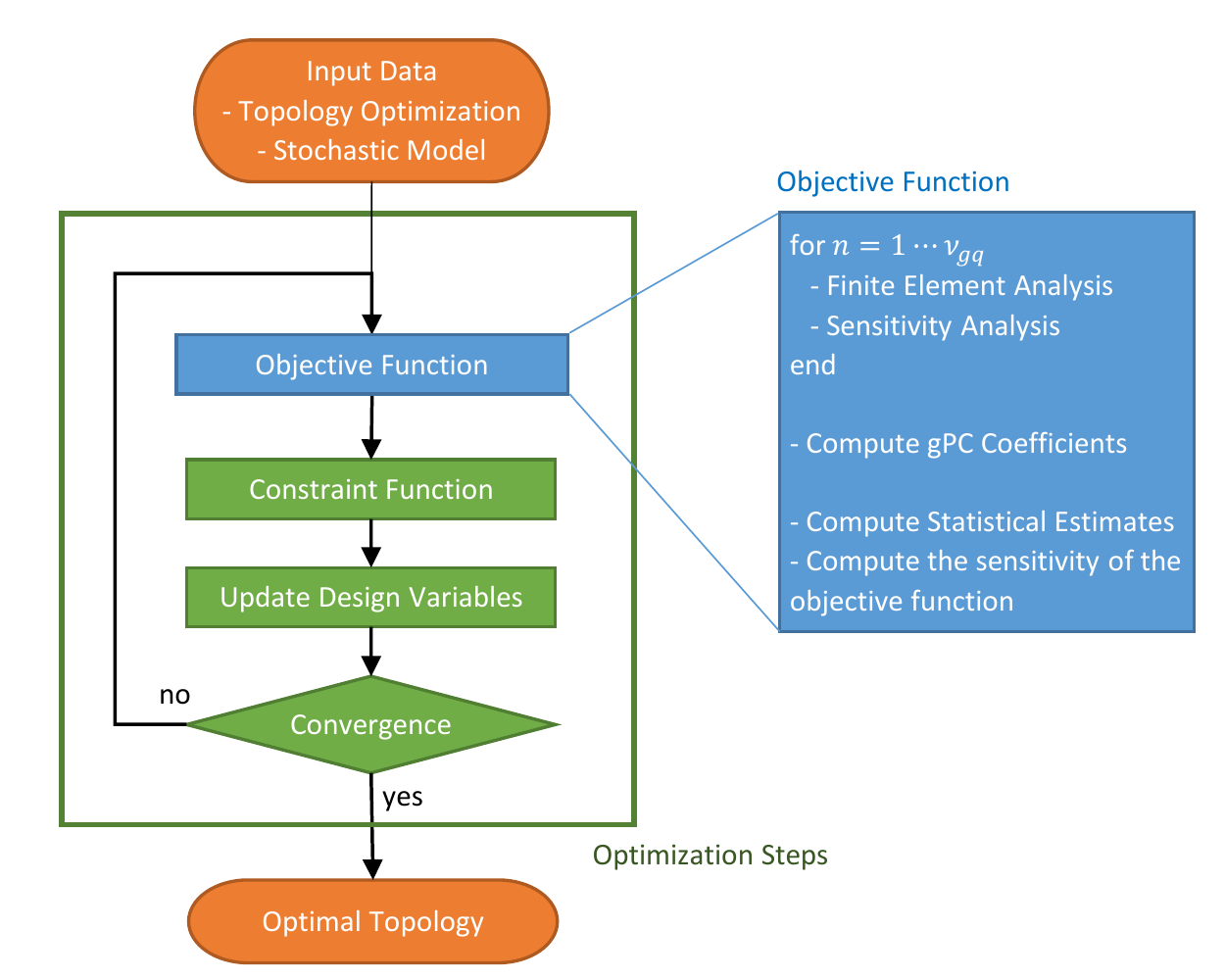}
\caption{General flow chart of the gPC RTO integrated procedure.}
\label{rto_flowchart_fig}
\end{figure*}

A flowchart of the proposed RTO algorithm is depicted in Figure~\ref{rto_flowchart_fig}.

\section{Numerical examples}
\label{num_examples}

The effectiveness of the proposed gPC RTO algorithm is addressed in this
section by means of a study that considers bidimensional mechanical systems
subjected to uncertain loads. The goal is to show that different statistical responses 
can be obtained when using the proposed gPC RTO design algorithm and a non-robust 
design strategy, where TO is done first (deterministically) and the propagation of uncertainties
is computed later, considering the deterministic optimized topology.

For the sake of accuracy verification, a reference crude Monte Carlo RTO solution 
is employed. This comparison allows one the verify the accuracy of the statistical measures 
obtained with the proposed gPC approach. The influence of the weight factor in the robust 
design is also addressed, as well as the different effects that are observed when a random load is treated as a random variable or a random field.


For the examples presented in the following section, consistent units are used.

\subsection{Cantilever beam design}

\begin{figure*}[h]
\centering
\includegraphics[scale=0.8]{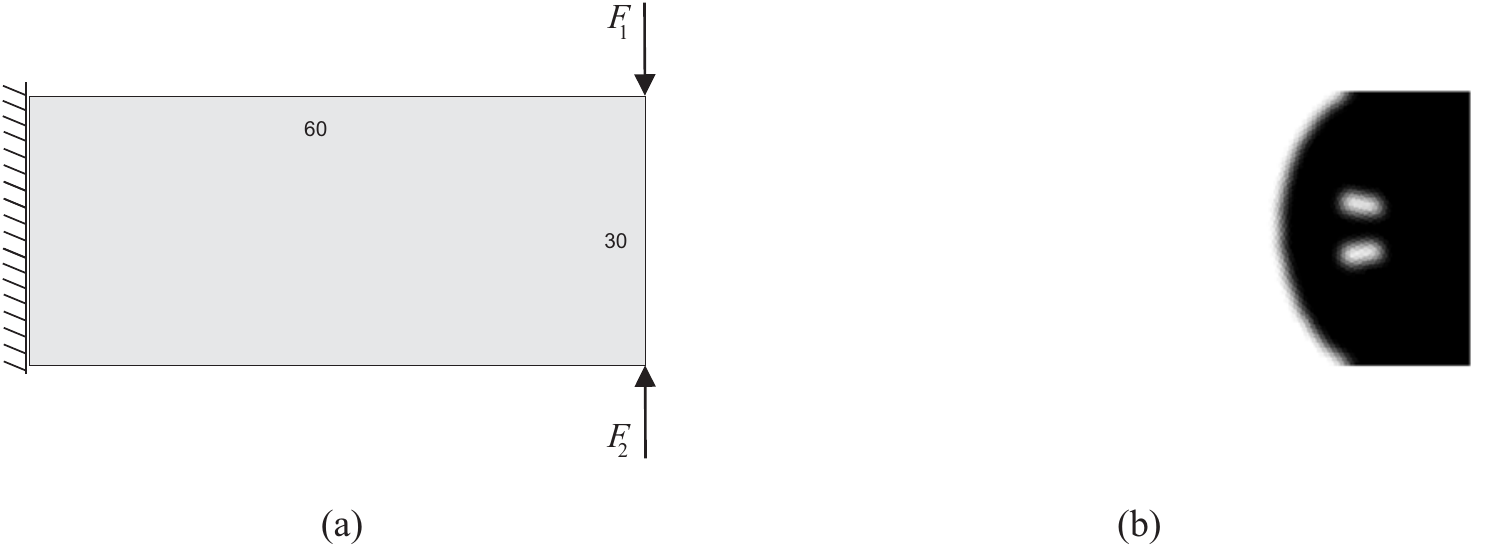}
\caption{Cantilever beam structure: (a) original configuration, (b) non-robust TO design.}
\label{cantilever_fig}
\end{figure*}

This first example consists of a simple cantilever beam subjected to a pair of vertical loads,
with uncertain magnitudes, applied at the two right edge corners, 
as illustrated in Figure~\ref{cantilever_fig}(a). This problem has been studied by Wu et al.\cite{WU201636}.

The vertical and horizontal dimensions of the structure 
are 30 and 60 units of length, respectively. The structure is composed of an isotropic 
material with Young modulus $E_0=1$ and Poisson ratio $\nu = 0.3$.
For the void material an elastic modulus value equal to $E_{min}=10^{-9}$ is 
employed. The prescribed volume fraction of material is set as $0.3$, the filter 
radius as $1.5$, the penalization factor $3$, and the design domain 
is discretized by means of a polygonal mesh with $N=7,200$ finite elements. 
The nominal (deterministic) configuration for this problem adopts the magnitude
of the two vertical forces as $F_1=F_2=1$, respectively.

On the other hand, in the stochastic case, magnitudes of the forces are assumed to be uncertain
and modeled by independent random variables $\SSpt \in \SS \mapsto \randvar{F}_1(\SSpt) \in \R$ and 
$\SSpt \in \SS \mapsto \randvar{F}_2(\SSpt) \in \R$, 
both defined on a suitable probability space $(\SS, \SF, \PM)$.
For the sake of simplicity, but being consistent with the physics of the mechanical problem, 
it is assumed that these two random variables are uniformly distributed on the same positive support
$\supp{F} = [F_{min}, F_{max}] \subset (0,+\infty)$. Three numerical studies are conducted in this example, 
where $\supp{F}$ is chosen as $[F_{min}, F_{max}] = [0.95, 1.05]$, $[F_{min}, F_{max}] = [0.9, 1.1]$ 
and $[F_{min}, F_{max}] = [0.8, 1.2]$. Note that these intervals correspond to symmetrical
variabilities of up to 5\%, 10\% and 20\% around the mean values $\mean{F_1}=\mean{F_2}=1$, 
respectively.

The non-robust TO, obtained using the \texttt{PolyTop} with MMA optimizer, 
is shown in Figure~\ref{cantilever_fig}(b). The lack of material on the left side of the cantilever 
is due to the two forces of equal magnitudes applied in opposite directions. Therefore, 
the stress in the cantilever is distributed only on the right side of the domain. However, 
to avoid instability (displacements going to infinity), a minimum value of elastic modulus
$E_{min}$ is used.

In order to perform the gPC RTO one needs to define the germ $\bm{\xi} = (F_1,F_2)$,
which is  over the region $[F_{min}, F_{max}] \times [F_{min}, F_{max}] \subset (0,+\infty) \times (0,+\infty)$, so that $\nu_{rv}=2$. 
In this case the optimal base for gPC expansion is given by the Legendre polynomials \cite{xiu2010}.
For the three different types of uniform distribution considered, a weight factor $w=1$ is employed, 
together with an expansion of order $p_{pc}=5$ (so that $1+\nu_{pc} = 21$)(this value was heuristically chosen to ensure the stochastic 
convergence.). A total number of $\nu_{gq} = 36$ collocation points 
is used to generate realizations of $\bm{\xi} = (F_1,F_2)$. To check the accuracy of the gPC RTO 
strategy, the same problem is addressed using the MC simulation with $\nu_{mc} = 10^4$ 
scenarios of loading magnitudes, a reference result dubbed MC RTO.

\begin{figure*}[h]
	\centering
	\includegraphics[scale=0.35]{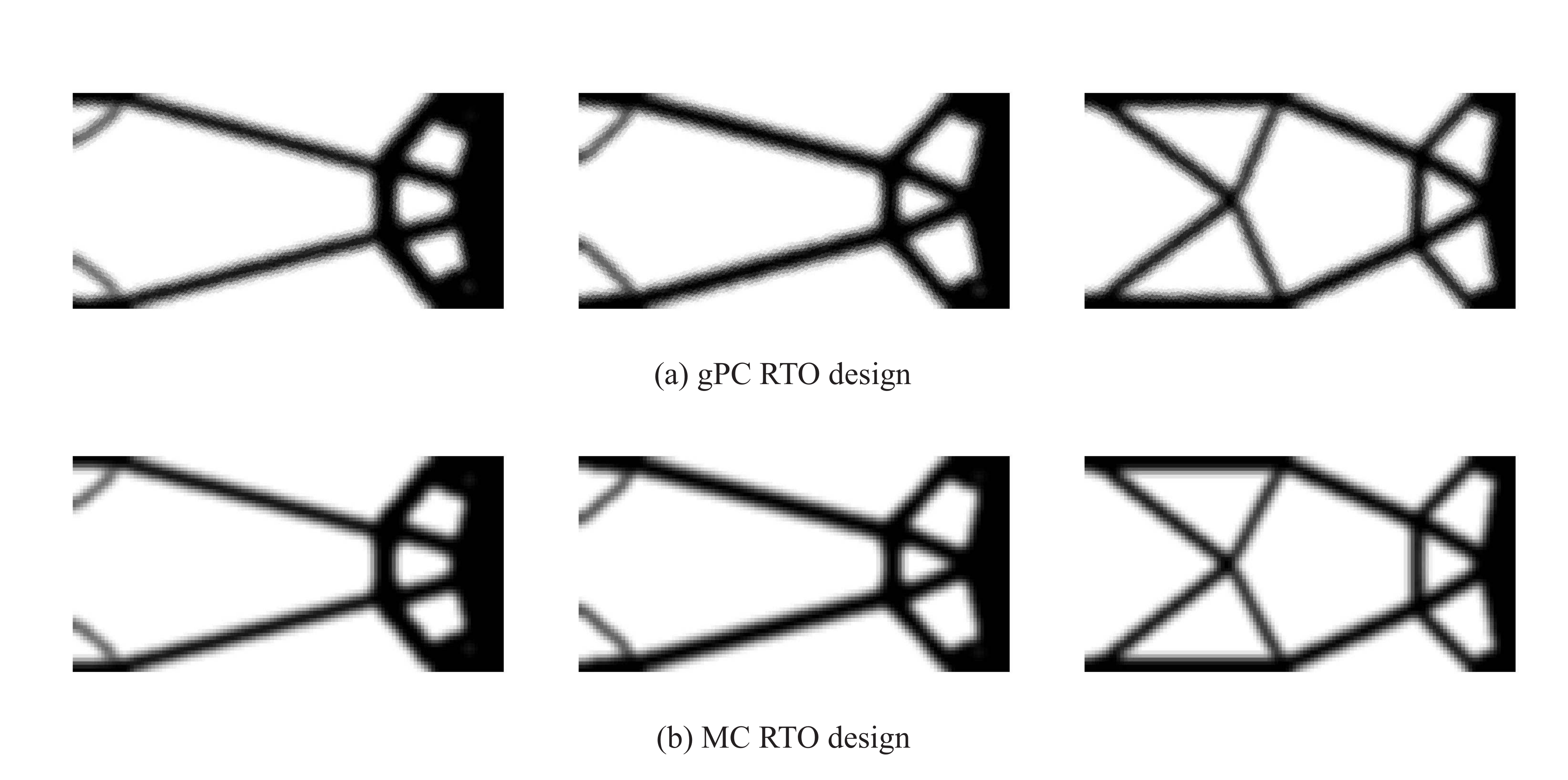}
	\caption{Optimized topologies for the cantilever beam 
		using gPC RTO and MC RTO designs, 
		for uniform distributions over the intervals 
		$[F_{min}, F_{max}] = [0.95,1.05]$ (left), 
		$[F_{min}, F_{max}] = [0.9,1.1]$ (center) and 
		$[F_{min}, F_{max}] = [0.8,1.2]$ (right).}
	\label{cantilever_gpc_fig}
\end{figure*}

In Figure~\ref{cantilever_gpc_fig} the reader can see the optimum topologies 
obtained by gPC RTO (top) and MC RTO (bottom), for different support of the 
random variable $\randvar{F}$.
The topologies shown in Figure~\ref{cantilever_gpc_fig} are different from 
the deterministic counterparts in Figure~\ref{cantilever_fig}(b) --
some extra members can be observed on the left side of the structure --
for different levels of uncertainties (length of $\supp{F}$). 
As the level of uncertainty increases, more members appear in the final 
topology. Finally, the robust designs using MC simulation present  equivalent topologies and statistical measures, which demonstrates the 
accuracy of the proposed gPC RTO approach.

Table~\ref{example1_tab} compares statistical estimates for the compliance 
expected value $\mu_{C}$ and standard deviation $\sigma_{C}$, in the cases
of robust and non-robust design.
Remember that, in this context, non-robust design means first optimizing the topology 
via classical (deterministic) TO and then using MC simulation to propagate 
the loading uncertainties through the mechanical system. A good agreement 
between robust strategies based on gPC and MC is noted, as well as that 
the statistical measures for the non-robust design tend to approach infinity, 
since there is no connection between the left and right sides of the domain.
It is also worth noting that, while the MC RTO needs $10^4$ evaluations of
the compliance function, the gPC RTO only needs $36$ evaluations.
This difference of three orders of magnitude demonstrates the efficiency
of the gPC RTO implementation.

\begin{table}[h!]
	\centering
	\caption{Low-order statistics of cantilever beam compliance for robust and non-robust TO strategies.}
	\vspace{5mm}
	\begin{tabular}{ccccccc}
		\toprule
		$\supp{F}$	& 	\multicolumn{2}{c}{gPC RTO} & \multicolumn{2}{c}{MC RTO} & \multicolumn{2}{c}{Non-robust TO}\\
		\midrule
		 	& 	$\mu_{C}$ & $\sigma_{C}$  & $\mu_{C}$ & $\sigma_{C}$  & $\mu_{C}$ & $\sigma_{C}$\\
		\cmidrule(r){2-3} \cmidrule(r){4-5} \cmidrule(r){6-7}
		$[0.95, 1.05]$ & 21.4 & 1.2 & 21.4	& 1.2 & 5.7 E7 & 6.7 E7\\
		$[0.90, 1.10]$ & 23.5 & 2.9 & 23.4	& 2.8 & 2.3 E8 & 2.7 E8\\
		$[0.80, 1.20]$ & 29.4 & 7.7 & 29.4	& 7.6 & 9.1 E8 & 1.1 E9\\	
		\bottomrule
	\end{tabular}
	\label{example1_tab}
\end{table}

\subsection{Michell type structure}

\begin{figure*}[h]
\centering
\includegraphics[scale=0.9]{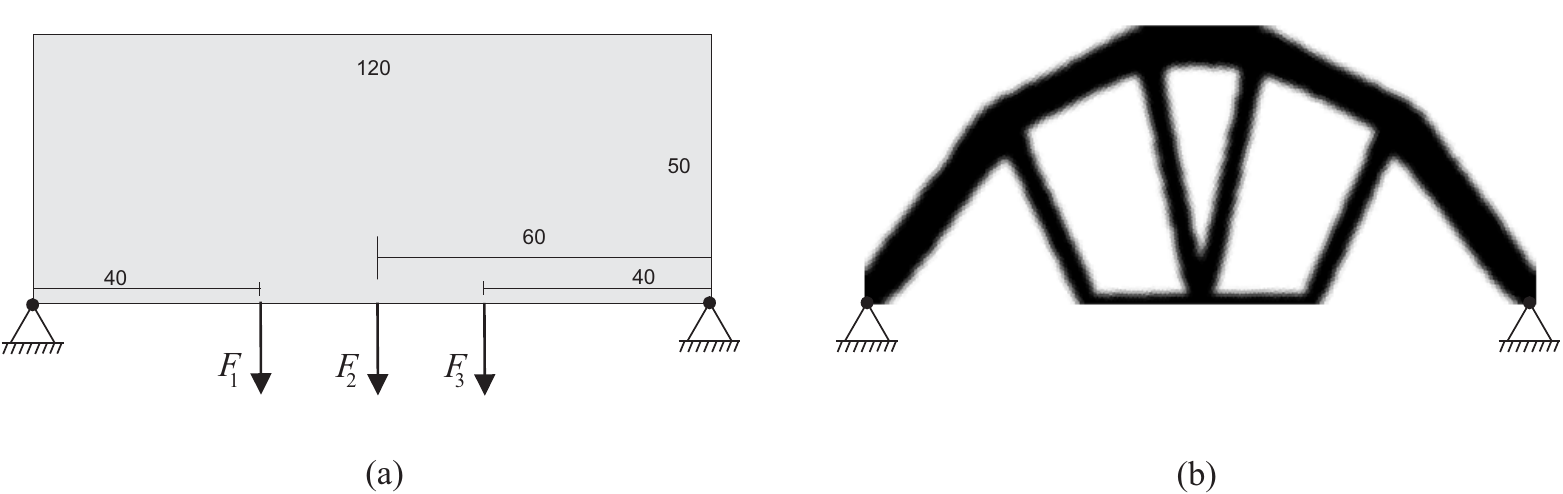}
\caption{Michell type structure: (a) original configuration, (b) non-robust TO design.}
\label{michell_fig}
\end{figure*}

In this second example RTO is performed on a simple Michell type structure
considering three loads, with uncertain directions, applied at the bottom edge
of the two dimensional system, as illustrated in Figure~\ref{michell_fig}(a). 

The length and height of the structure are equal to 120 and 50 units, respectively. 
The design domain is discretized with a polygonal mesh with $N = 12,000$ 
finite elements, and all other parameters are the same as in the first example. 
For the deterministic case, magnitudes and directions of the three forces 
are defined as $F_1=1$, $F_2=2$, $F_3=1$, and $\alpha_1=\alpha_2=\alpha_3=-90º$,
respectively.

\begin{figure*}[h]
\centering
\includegraphics[scale=0.5]{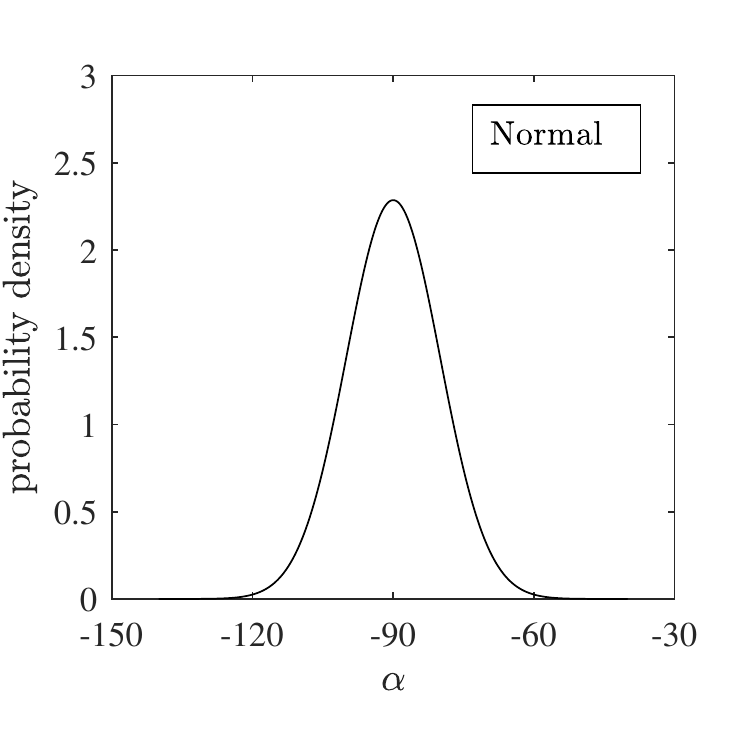}
\includegraphics[scale=0.5]{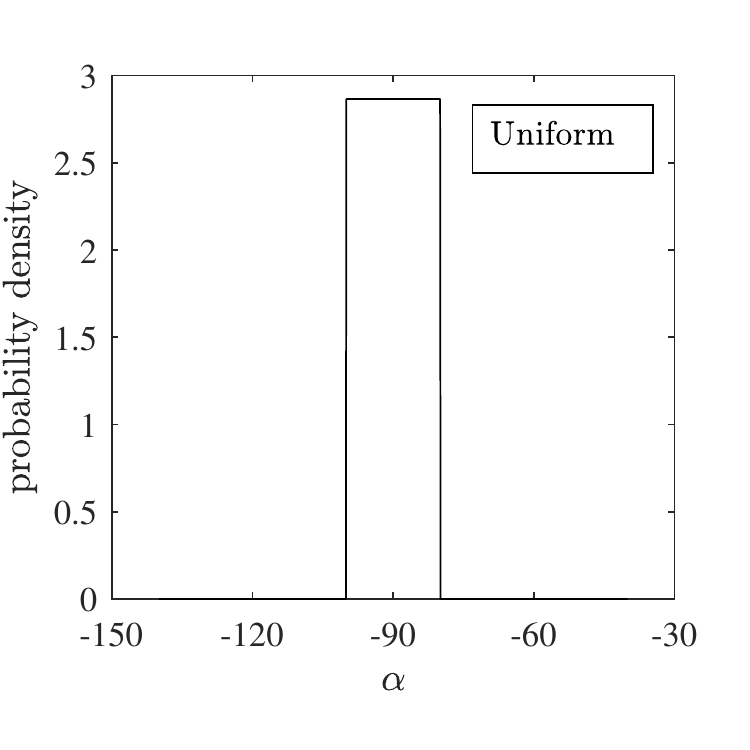}
\includegraphics[scale=0.5]{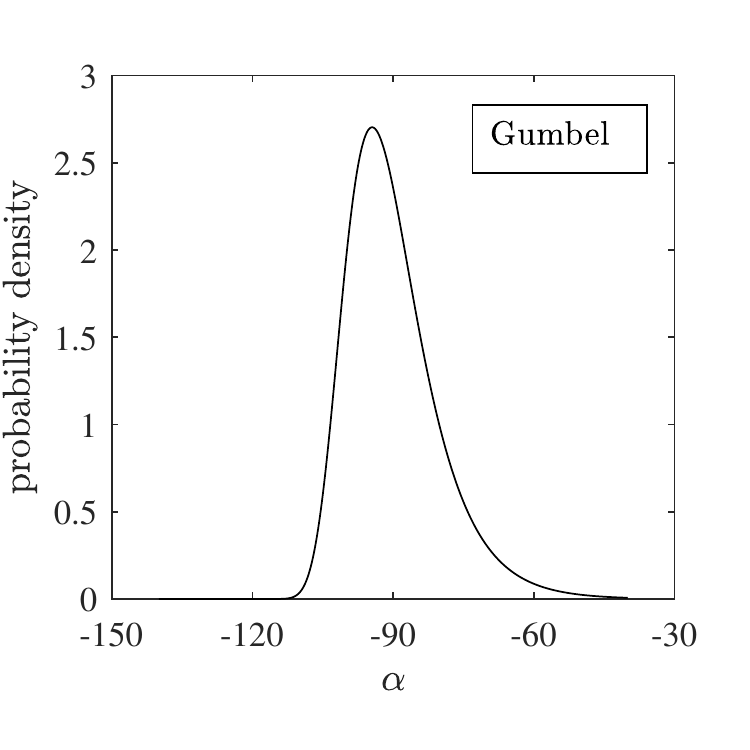}
\caption{Probability distributions for the loads angles: normal (left), uniform (center) and Gumbel (right).}
\label{michell_pdf_fig}
\end{figure*}

Meanwhile, on the stochastic case, the forces directions are modeled as the 
independent and identically distributed random variables $\SSpt \in \SS \mapsto \randvar{A}_1(\SSpt) \in \R$,
$\SSpt \in \SS \mapsto \randvar{A}_2(\SSpt) \in \R$ and
$\SSpt \in \SS \mapsto \randvar{A}_3(\SSpt) \in \R$.
Three scenarios of probabilistic distribution are analyzed:
(i) Normal, (ii) Uniform, and (iii) Gumbel.
For the Normal and Gumbel distributions, mean values are assumed
to be equal to the nominal values of $\alpha_1$, $\alpha_2$ and $\alpha_3$,
with all the standard deviations equal to $10º$. In the Uniform case,
the three supports are defined by the interval $[A_{min},A_{max}] = [-100º,-80º]$.
The probability density functions of these distributions are illustrated in 
Figure~\ref{michell_pdf_fig}.

\begin{figure*}
\centering
\subfigure[Normal]{\includegraphics[scale=0.55]{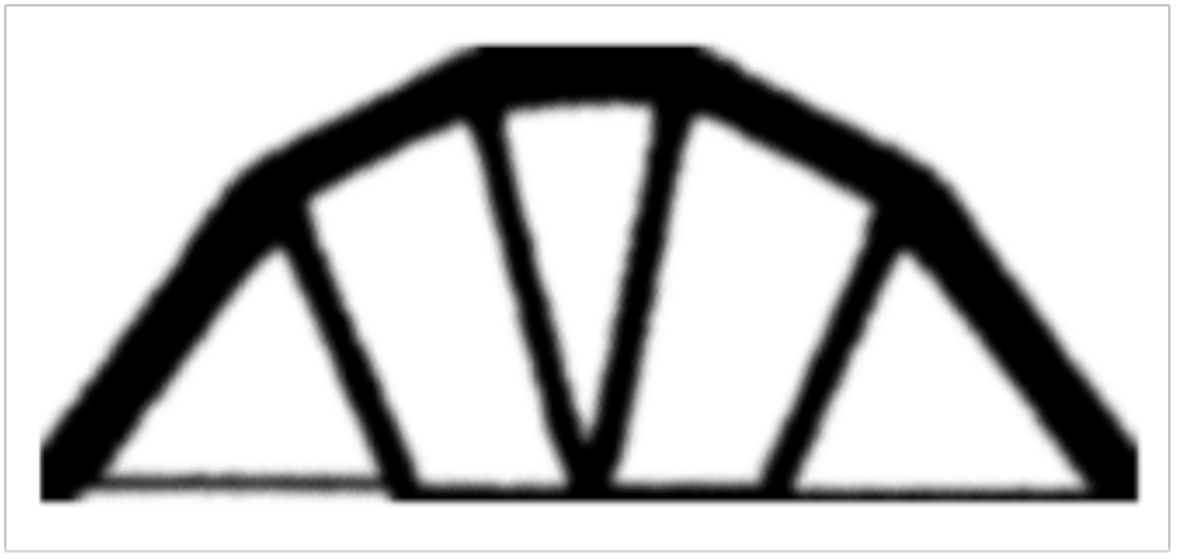}}
\subfigure[Uniform]{\includegraphics[scale=0.55]{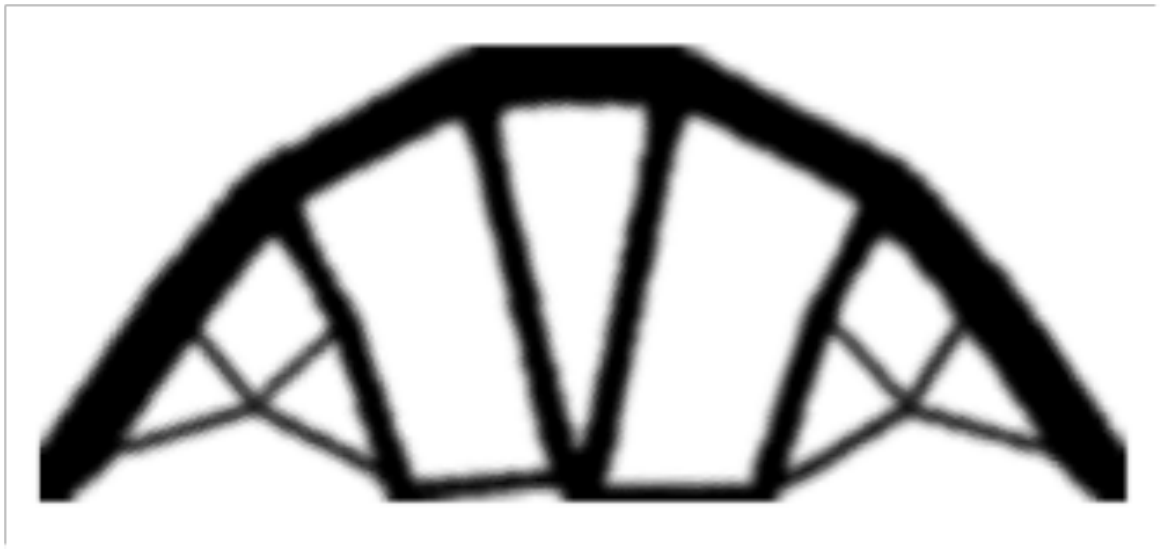}}\\
\subfigure[Gumbel]{\includegraphics[scale=0.55]{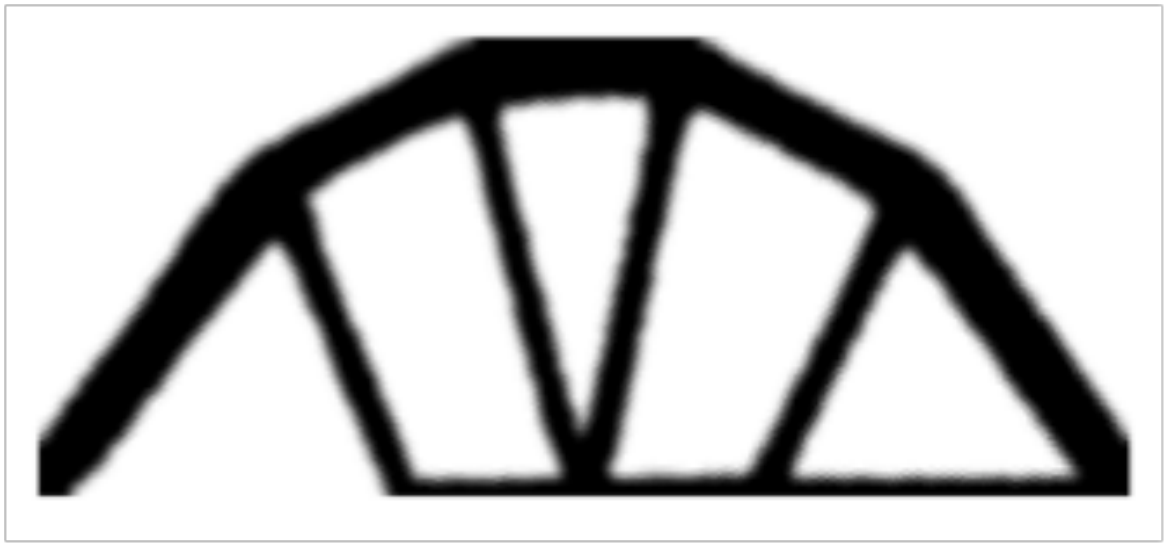}}
\caption{Optimized topologies for the Michell type structure using gPC RTO design,
with different probability distributions for load angle: (a) Normal, (b) Uniform and (c) Gumbel.}
\label{michell_gpc_fig}
\end{figure*}

Now the germ is defined as $\bm{\xi} = (\randvar{A}_1,\randvar{A}_2,\randvar{A}_3)$,
thus $\nu_{rv}=3$, and the family of orthogonal polynomials (basis) is chosen according 
to the germ support. For simplicity, Hermite polynomials are used in the case of Gaussian 
or Gumbel parameters, while Legendre polynomials are the option when the germ is uniform distributed.
Employing gPC RTO with an expansion of order $p_{pc}=5$ 
(thus $1+\nu_{pc} = 56$) total number of $\nu_{gq}=216$ collocation points and and
weight factor value $w=1$, one obtains the robust designs shown 
in Figure~\ref{michell_gpc_fig}, 
where connections at fixed points are created to balance the horizontal components 
of non-vertical forces. Note that the non-robust design in Figure~\ref{michell_fig}(b) 
only presents four bars connected at the forces application points, no connection at the 
joints can be seen. This occurs because the forces are always vertical. 
However, when gPC RTO design is used, there are connections at the joints, 
because the angle variability induces horizontal force components. 

\begin{figure}
\centering
\subfigure[Normal]{\includegraphics[scale=0.35]{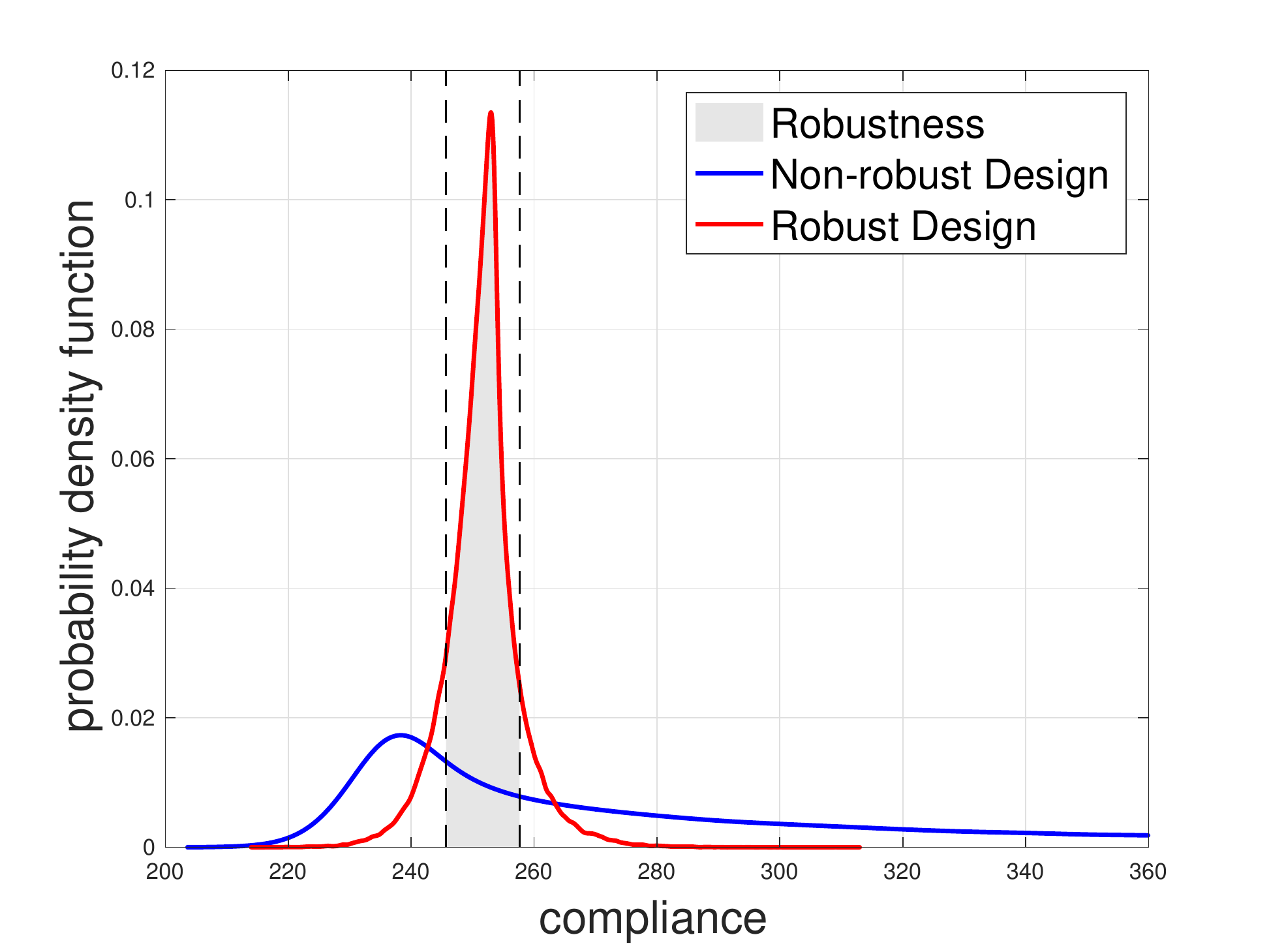}}
\subfigure[Uniform]{\includegraphics[scale=0.35]{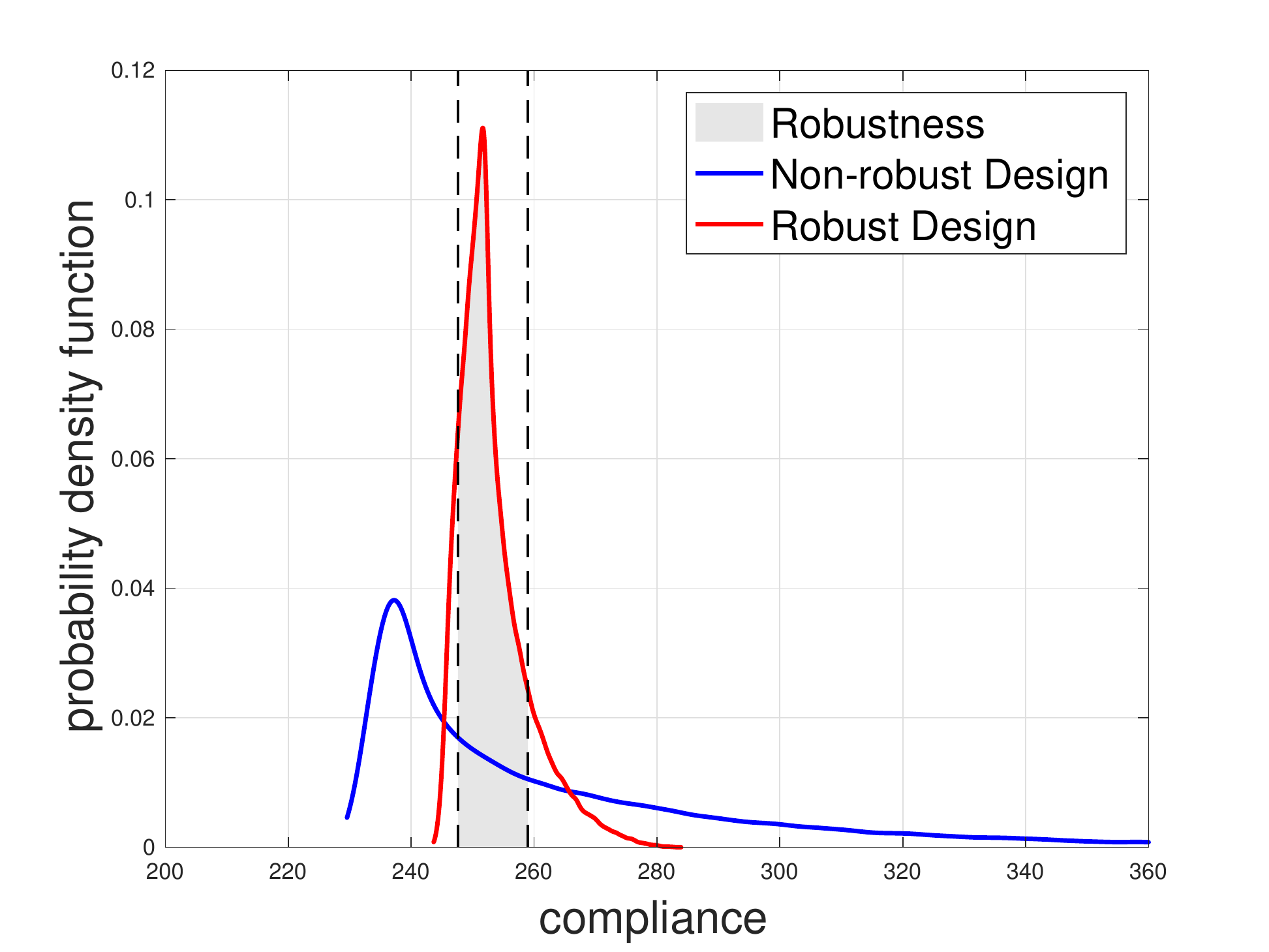}}\\
\subfigure[Gumbel]{\includegraphics[scale=0.35]{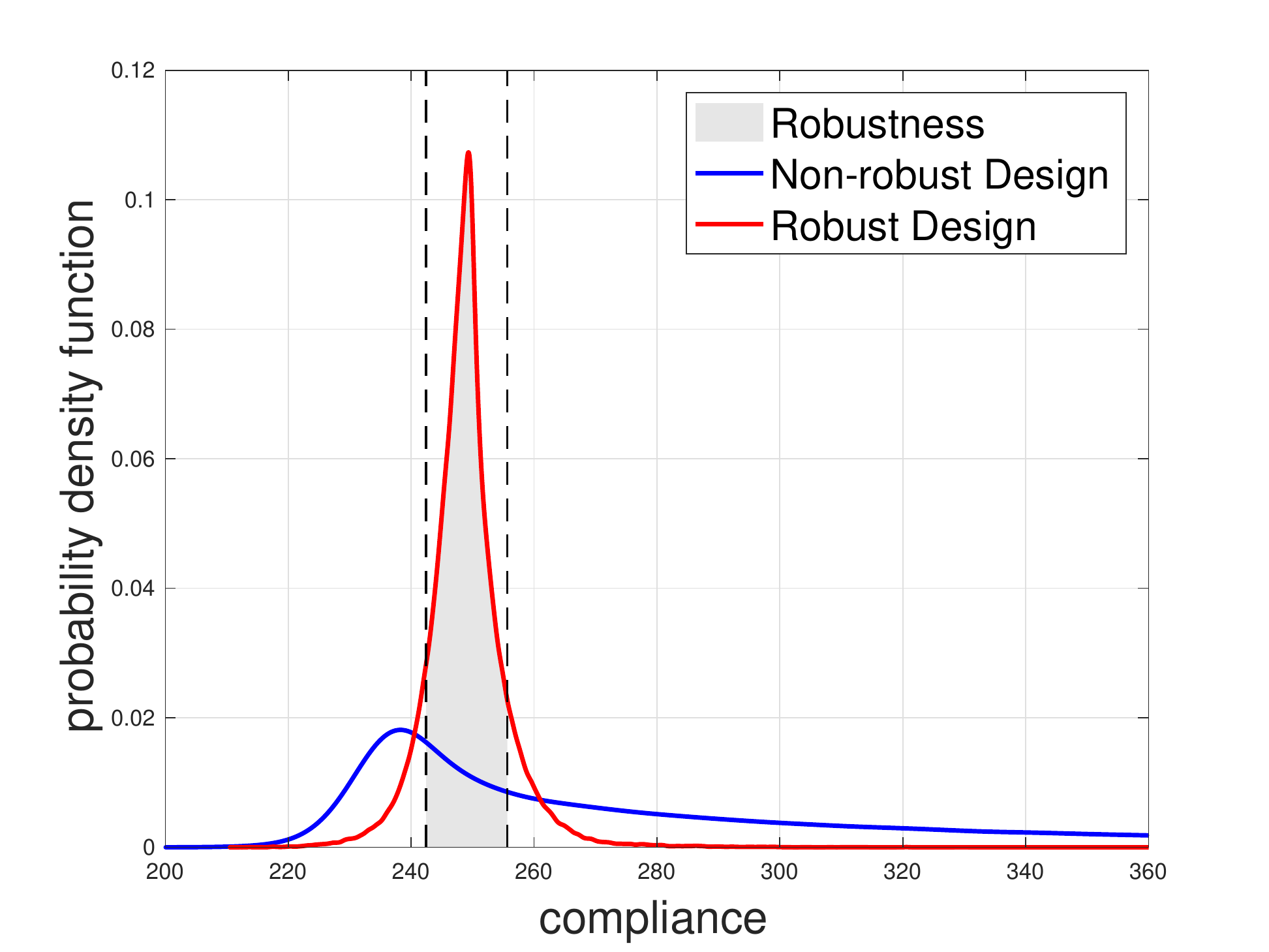}}
\caption{Probability density function of the compliance for the Michell type structure using non-robust
and gPC RTO robust design, with different probability distributions for load angle: 
(a) Normal, (b) Uniform and (c) Gumbel.}
\label{michell_Cpdf_fig}
\end{figure}

The statistical results of the robust design compared with the non-robust counterpart 
can be appreciated in Figure~\ref{michell_Cpdf_fig} and Table~\ref{example2_tab},
which show the compliance probability densities and their low order statistics, respectively, 
for the different distributions considered in the force angle.
One can observe from these simulation results that the range variability of compliance is reduced, which implies that the robust design is less sensitive to loading uncertainties than its non-robust counterpart.
As shown in Figure~\ref{michell_gpc_fig}, the final topologies are symmetric for the normal and 
uniform distributions but is asymmetric for the Gumbel distribution.

\begin{table}[h!]
	\centering
	\caption{Low-order statistics of the compliance for the Michell type structure using robust and non-robust TO strategies.}
	\vspace{5mm}
	\begin{tabular}{lcccc}
		\toprule
		distribution & \multicolumn{2}{c}{gPC RTO} & \multicolumn{2}{c}{Non-robust TO}\\
		\midrule
		 	& 	$\mu_{C}$ & $\sigma_{C}$  & $\mu_{C}$ & $\sigma_{C}$\\
		\cmidrule(r){2-3} \cmidrule(r){4-5}
		Normal  & 251.6 & 6.0 & 314.2 & 113.3\\
		Uniform & 253.3 & 5.7 & 262.7 & 33.3\\
		Gumbel  & 249.1 & 6.7 & 312.5 & 128.9\\
		\bottomrule
	\end{tabular}
	\label{example2_tab}
\end{table}

\begin{figure*}
\centering
\subfigure[$w=0$]{
\includegraphics[scale=0.6]{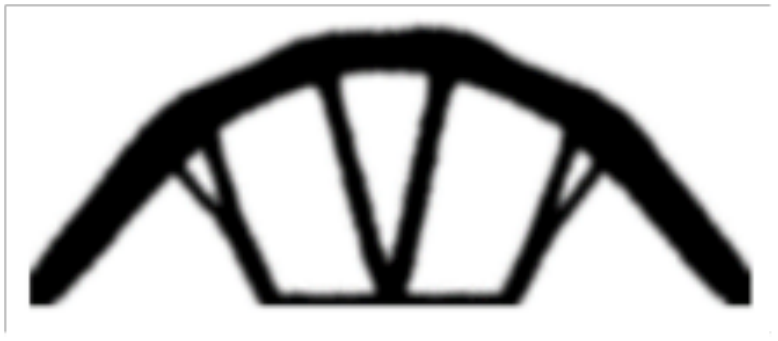}
\includegraphics[scale=0.6]{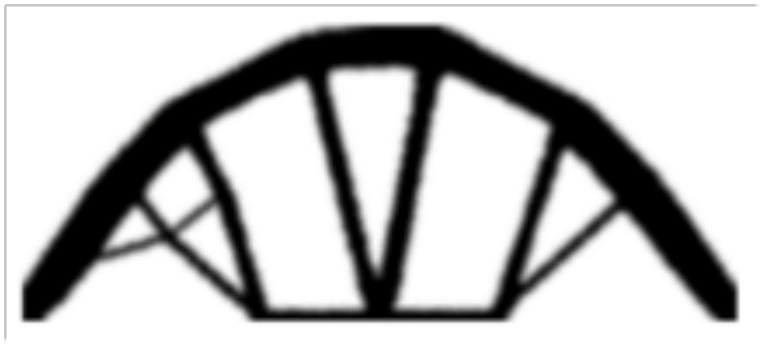}
\includegraphics[scale=0.6]{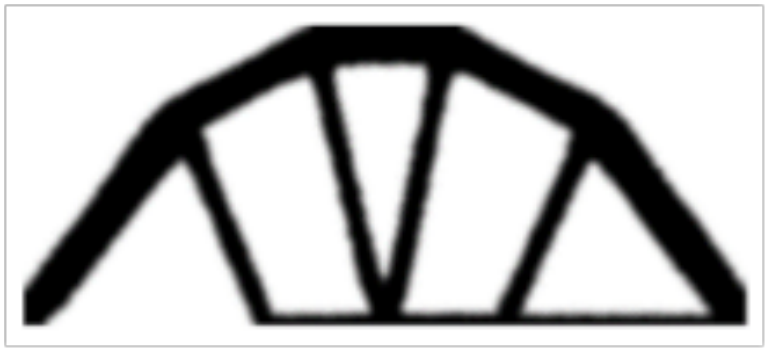}}\\
\subfigure[$w=1$]{
\includegraphics[scale=0.6]{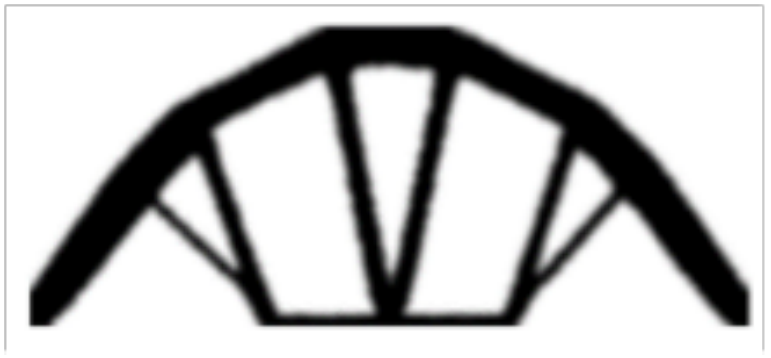}
\includegraphics[scale=0.6]{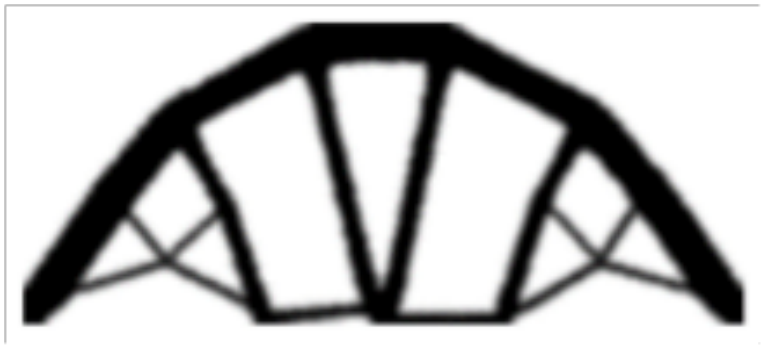}
\includegraphics[scale=0.6]{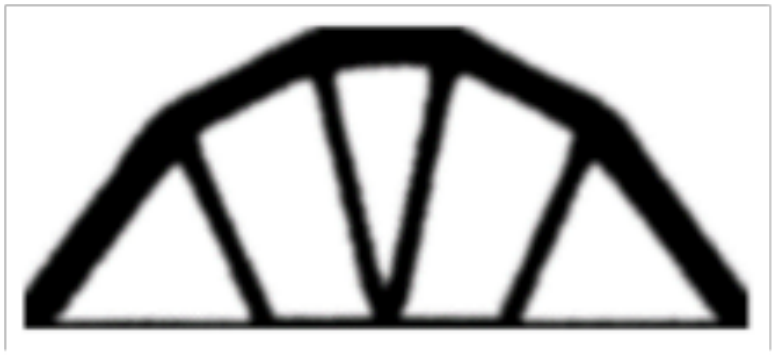}}\\
\subfigure[$w=2$]{
\includegraphics[scale=0.6]{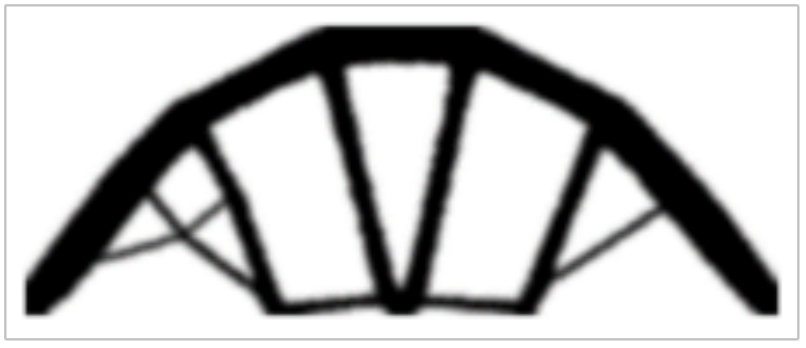}
\includegraphics[scale=0.6]{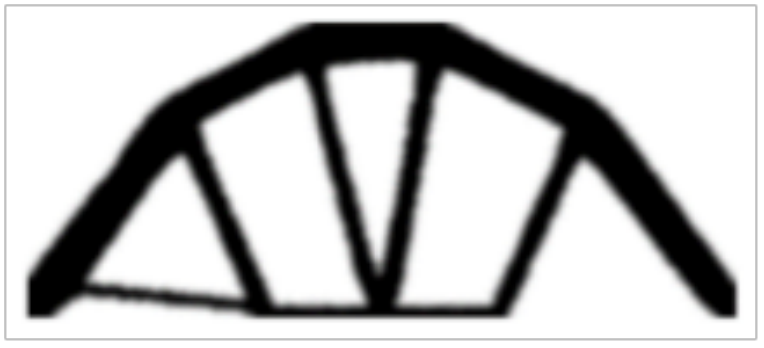}
\includegraphics[scale=0.6]{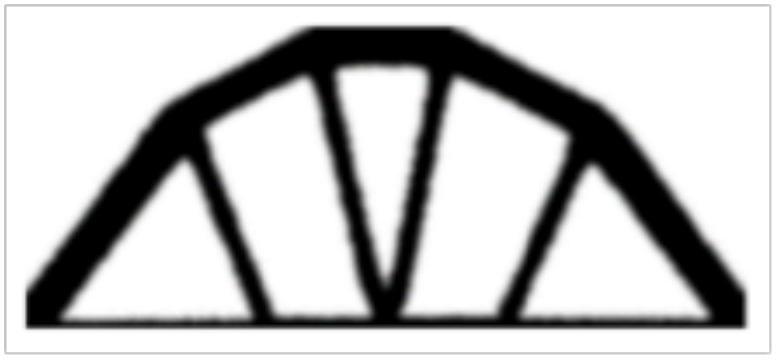}}\\
\subfigure[$w=3$]{
\includegraphics[scale=0.6]{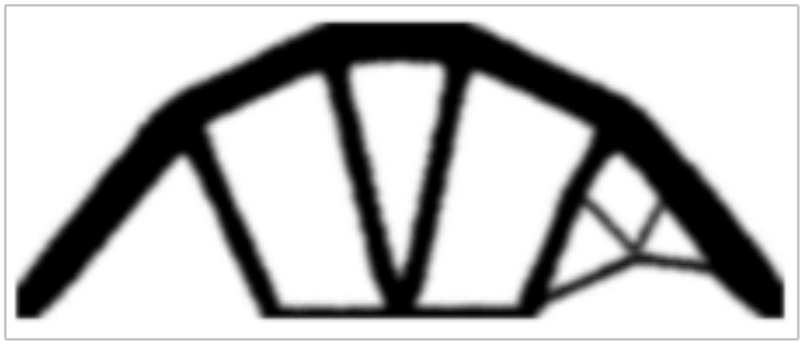}
\includegraphics[scale=0.6]{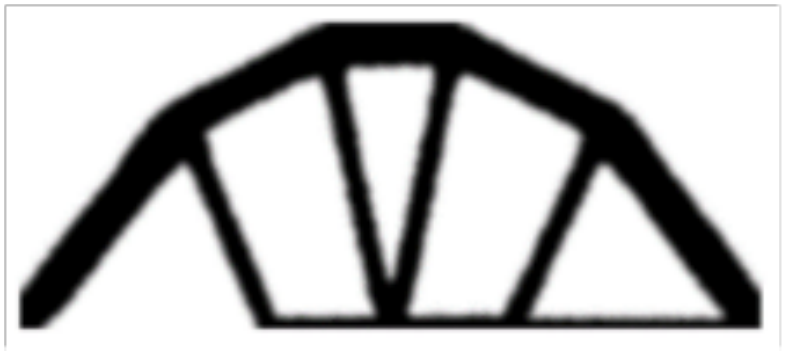}
\includegraphics[scale=0.6]{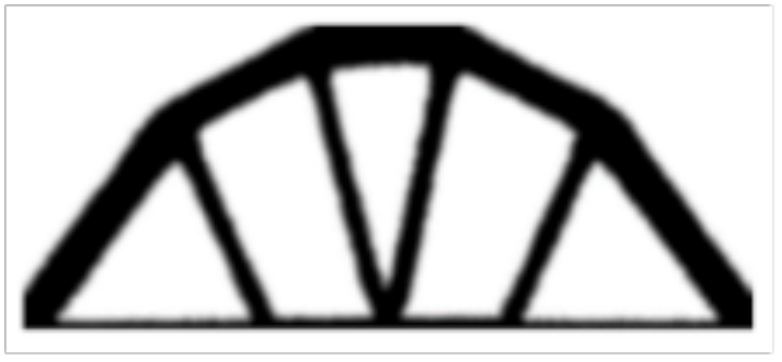}}
\caption{Optimized topologies for the Michell type structure using gPC RTO strategy,
for different values of weight $w$ and different uniform distributions for the angle:
$[-95º,-85º]$ (left), $[-100º,-80º]$ (middle) and $[-110º,-70º]$ (right).}
\label{michell_k_fig}
\end{figure*}

In the second analysis for this example, the influence of the weighting factor 
in the robust optimization process is addressed. 
The aim is to minimize the variability by increasing the value of $w$, because this 
factor is directly related to the standard deviation term on the objective function. 

Three different uniform distributions are considered for the random angle
of the force and their supports are respectively defined by the intervals
$[A_{min},A_{max}] = [-95º,-85º]$, $[A_{min},A_{max}] = [-100º,-80º]$ and \linebreak
$[A_{min},A_{max}] = [-110º,-70º]$. The gPC RTO design strategy is employed 
for $w \in \{0,1,2,3\}$, generating the optimal topologies shown in 
Figure~\ref{michell_k_fig}.
According to Table~\ref{example2_tab2}, all the designs shown in Figure~\ref{michell_k_fig} present significant lower expected compliance and standard deviation values when compared to the non-robust solution.
The highest values of standard deviation are obtained using $w=0$, since we are minimizing only the expected compliance. For $w>0$, both expected compliance and its standard deviation contribute to the objective function and we can observe that as the value of $w$ increases, the standard deviation decreases. Based on the numerical experiments presented in Table~\ref{example2_tab2}, we recommend the value $w=3$, for practical use, because it leads to the best values of standard deviation with only a slight change in the expected compliance values.


\begin{table*}[h!]
	\centering
	\caption{Low-order statistics of the compliance for the Michell type structure using gPC RTO,
	with different force angles distributions and different weight factors.}
	\vspace{5mm}
	\begin{tabular}{cccccccc}
		\toprule
		 & & \multicolumn{6}{c}{$\supp{A}$}\\
		 \cmidrule(r){3-8}
		 & & \multicolumn{2}{c}{$[-  95º,- 85º]$}  & \multicolumn{2}{c}{$[-100º, -80º]$}  & \multicolumn{2}{c}{$[-110º, -70º]$}\\
		 & w & $\mu_{C}$ & $\sigma_{C}$  & $\mu_{C}$ & $\sigma_{C}$  & $\mu_{C}$ & $\sigma_{C}$\\
		\cmidrule(r){3-4} \cmidrule(r){5-6} \cmidrule(r){7-8}
		gPC RTO & 0 & 241.8 &   5.2 & 256.0 & 10.6 & 249.0 & 10.2\\	
	    gPC RTO & 1 & 247.2 &   4.3 & 253.3 &   5.7 & 249.2 &   6.4\\
		gPC RTO & 2 & 253.0 &   3.1 & 249.7 &   4.0 & 250.0 &   6.0\\
	    gPC RTO & 3 & 248.9 &   2.4 & 247.0 &   2.9 & 250.1 &   5.6\\
		Non-robust TO & - & 364.4 & 14.2 & 366.1 & 28.4 & 373.0 & 56.9\\	
		\bottomrule
	\end{tabular}
	\label{example2_tab2}
\end{table*}

\subsection{2D bridge structure}

\begin{figure*}
	\centering
	\includegraphics[scale=0.335]{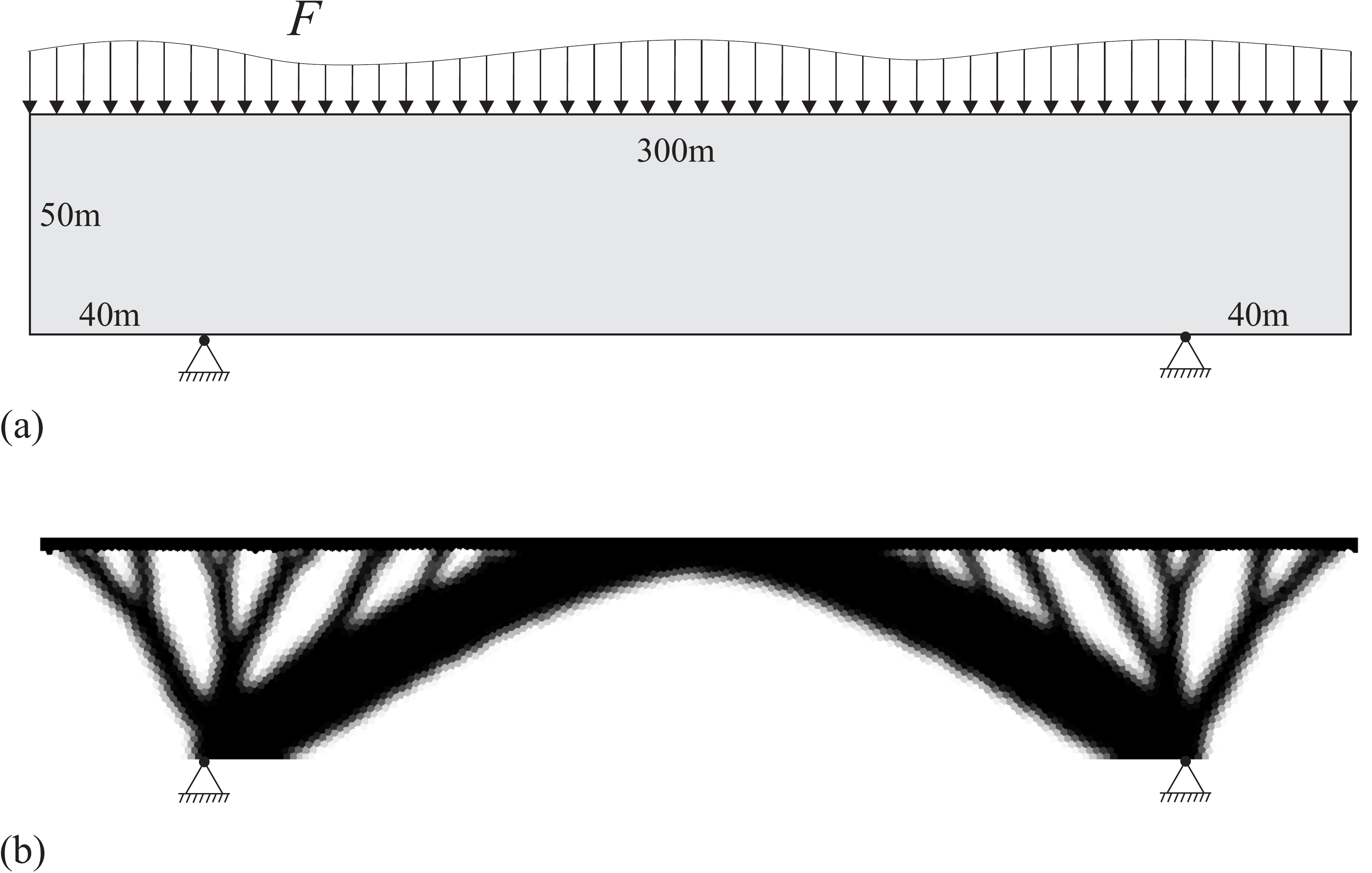}
	\caption{2D Bridge structure: (a) original configuration, (b) non-robust RO design.}
	\label{bridge_fig}
\end{figure*}

This last example corresponds to a simple bridge structure subjected to an uncertain distributed loading 
at the top edge, as illustrated in Figure~\ref{bridge_fig}(a). The nominal load is uniform throughout the structure, with magnitude per unit of length equal to $F=1$. 

For the stochastic case, the load magnitude per unit of length in each point is assumed to be a
Gaussian random field $(\SSpt,x) \in \SS \times [0,l] \mapsto \randvar{F}(\SSpt,x) \in \R$ with correlation function 
\begin{equation}
	\corr{\randproc{F}} (x,x') = \sigma_F,
\end{equation}
such that the loads at any pair of points $x$, $x'$ are fully correlated. The mean and standard deviation
of the random field $\randvar{F}(\SSpt,x)$ are assumed as $\mean{F}=1$ and $\stddev{F} = 0.3$, respectively.

For the optimization process, the prescribed volume fraction of material is set as 0.3, the filter radius is set as 3, 
the penalization factor 3, and the design domain is discretized with a polygonal mesh with $N=10,000$ finite elements.
The non-robust TO of the 2D bridge, performed using \texttt{PolyTop} with MMA optimizer, is shown in Figure~\ref{bridge_fig}(b). 
Furthermore, the first two rows of finite elements on the top of structure are fixed during the optimization process, to ensure that the 
bridge remains attached to the loading conditions.
Allowing the final results to be more realistic. It is observed in 
Figure~\ref{bridge_fig}(b) that the non-robust design leads to a final topology which is 
similar to the classical case of a 2D bridge under an uniformly distributed load.

For the purpose of numerical computation, the random field $\randvar{F}(\SSpt,x)$ is discretized
by means of $\bm{\xi}=(\randvar{F}_1)$, a single Gaussian random variable for which low-order statistics
are the same as for the random field. The gPC RTO design is obtained using an expansion 
of order $p_{pc}=5$ (with $1+\nu_{pc}=6$) and a total number of $\nu_{gq}=6$ collocation points for Hermite orthogonal polynomials, and
weight factor values $w \in \{0,1,2,3\}$.

\begin{figure}
	\centering
	\includegraphics[scale=0.57]{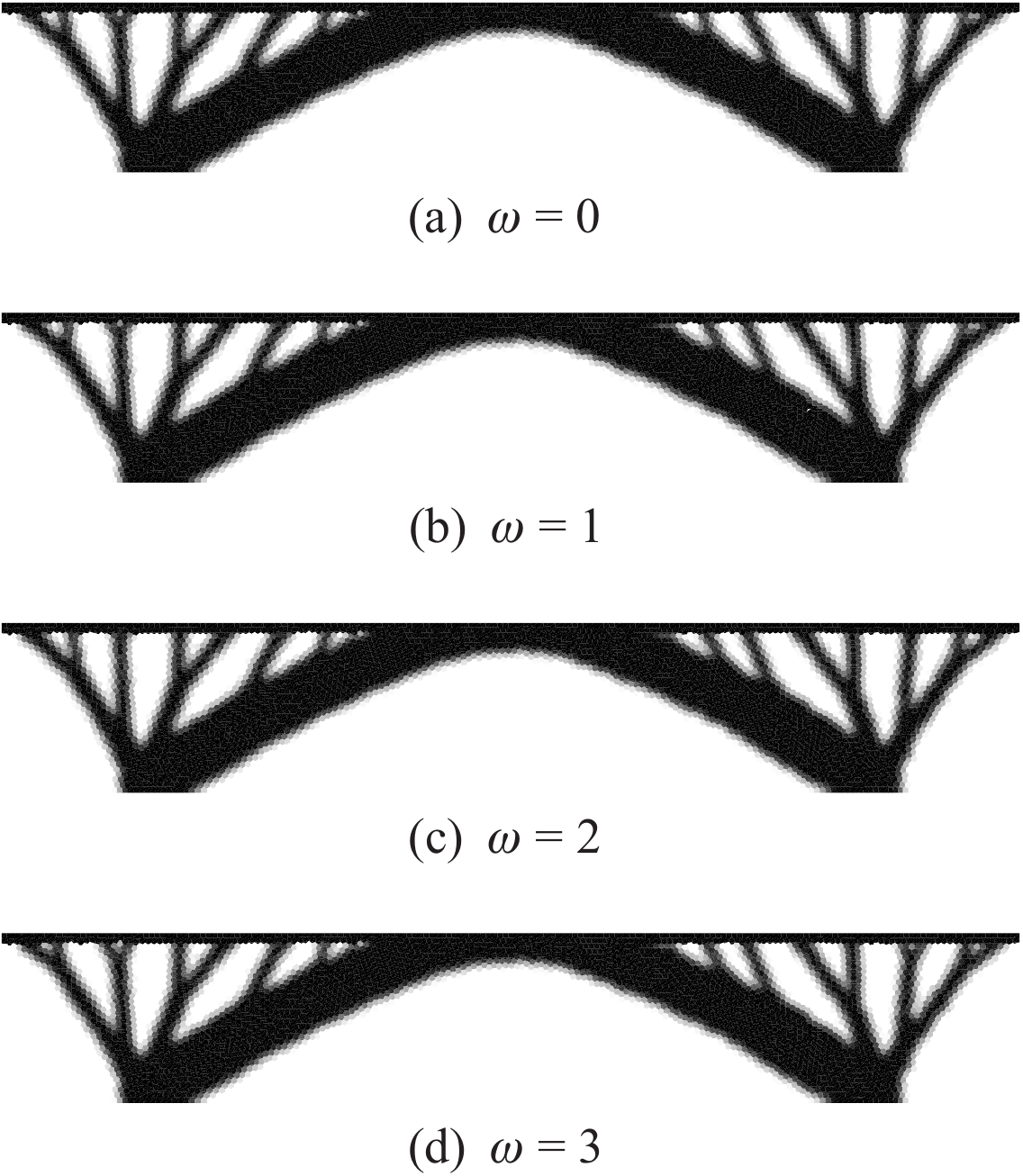}
\caption{Robust design for the 2D bridge structure with fully correlated distributed load,
	for different values of weight $w$.}
\label{bridge_k_fig}
\end{figure}

The final results are shown in Figure~\ref{bridge_k_fig},
where one can observe that some bars connected at the bottom of the bridge are different from those 
of the non-robust case in Figure~\ref{bridge_fig}(b). Moreover, as the value of $w$ increases, 
the structure presents a more robust physical form, which represents a consistent result, because the standard 
deviation of the compliance is being forced to be smaller. The corresponding mean value and 
standard deviation of the compliance function, for the different values of $w$ employed,
are given in Table~\ref{example3_tab}.

\begin{table}[h!]
	\centering
	\caption{Low-order statistics of the compliance for a 2D bridge using gPC RTO and different weight factors.}
	\vspace{5mm}
	\begin{tabular}{lcccc}
		\toprule
		 $w$	& 	$\mu_{C}$ & $\sigma_{C}$\\
		\midrule
		0  & 5.4443 E5 & 3.0666 E5\\
		1  & 5.4481 E5 & 3.0657 E5\\
		2  & 5.4462 E5 & 3.0646 E5\\
		3  & 5.4455 E5 & 3.0642 E5\\
		\bottomrule
	\end{tabular}
	\label{example3_tab}
\end{table}

As a second analysis, the random load $\randvar{F}(\SSpt,x)$ is assumed to have the same
low-order statistics as before, but an exponentially decaying correlation function 
\begin{equation}
	\corr{\randproc{F}} (x,x') = \stddev{\randvar{F}} \, \exp{\left( -\frac{|x-x'|}{l_{corr}} \right)},
\end{equation}
where $l_{corr}$ is a correlation length for the random field. Note that, by this assumption, the loads at
any two points $x$, $x'$ in the 2D bridge are partially correlated. 
If the correlation length is increased, a strong correlation is obtained between the points $x$, $x'$,
so that $l_{corr}=\infty$ implies a perfectly correlated random field --- the previous case where the field depends on a single random variable.
On the other hand, when $l_{corr}=0$, the random field is completely uncorrelated --- many independent random 
variables are necessary for an accurate computational representation. In order to avoid the two limit cases,
$l_{corr}=120$ is chosen.

In terms of computational representation for numerical calculations, the random field $\randvar{F}(\SSpt,x)$
is discretized with the aid of Karhunen–Lo\`{e}ve expansion described in section~\ref{kl_expansion}. The number of
terms in this expansion is chosen in a heuristic way, seeking to satisfy the criterion presented in (\ref{kl_conv_crit}).
A good compromise between accuracy and computational efficiency is obtained with $\nu_{kl}=7$. Therefore, the germ is
$\bm{\xi}=(\randvar{F}_1,\randvar{F}_2,\cdots,\randvar{F}_{\nu_{rv}})$, a set of $\nu_{rv}=7$ independent Gaussian 
random variables for which low-order statistics are the same as for the random field. Then, for this case we use a total number of $\nu_{gq}=279936$ collocation points.

\begin{figure}
\centering
\includegraphics[scale=0.4]{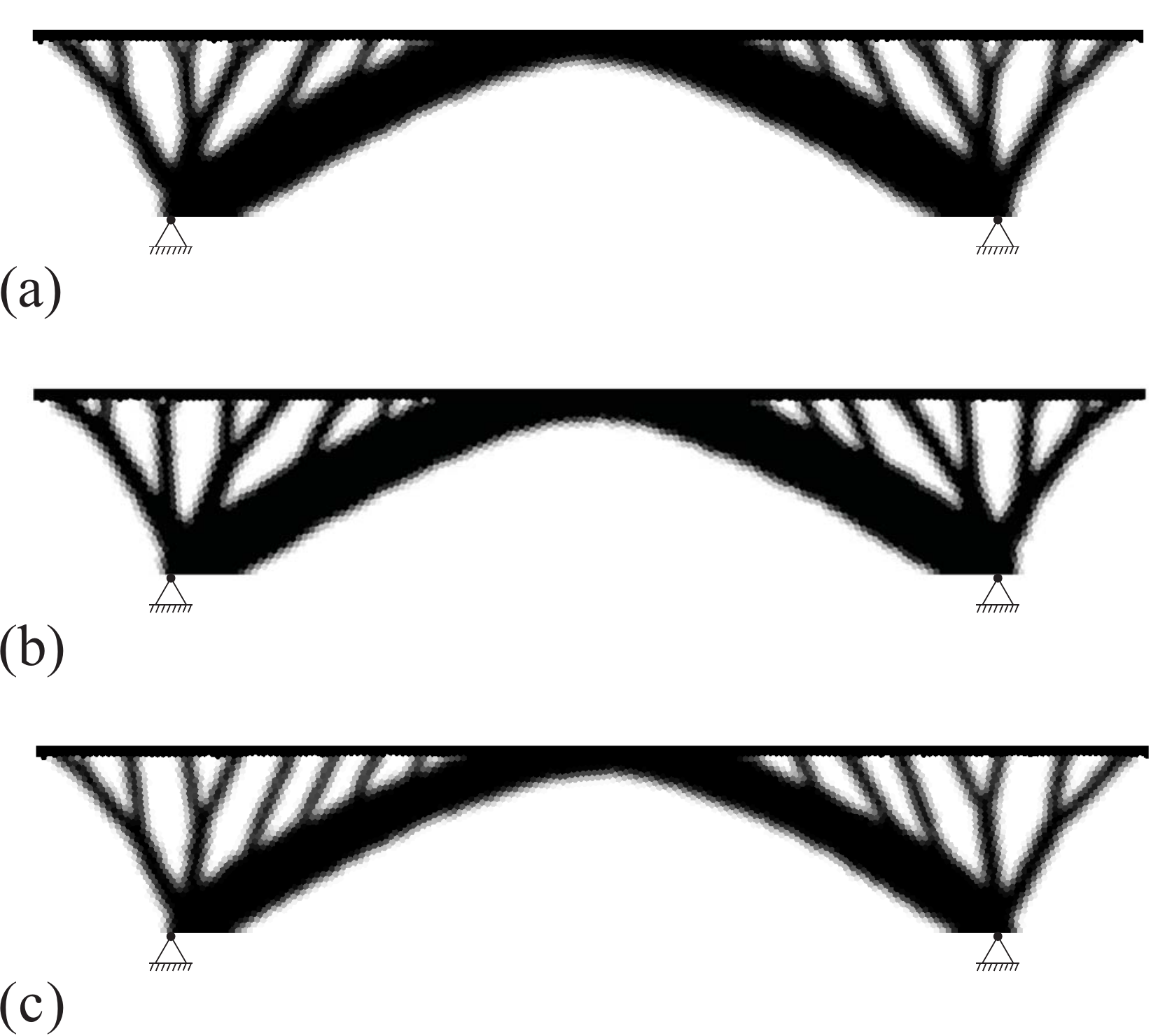}
\caption{Optimized topologies for the 2D bridge structure:
	(a) non-robust TO with fully correlated load, (b) gPC RTO with fully correlated load, (c) gPC RTO with partially correlated load.}
\label{bridge_gpc_fig}
\end{figure}

A comparison between non-robust and gPC RTO design, for $w=1$ and the different types of
distributed load considered, are shown in Figure~\ref{bridge_gpc_fig}. The difference between
the three obtained topologies is very clear, and can also be appreciated in Table~\ref{example3_tab2},
which shows the low-order statistics of the compliance in all cases analyzed.

\begin{table*}[h!]
	\centering
	\caption{Low-order statistics of the compliance for a 2D bridge considering different design scenarios.}
	\vspace{5mm}
	\begin{tabular}{cccccc}
		\toprule
		\multicolumn{2}{c}{Non-robust TO} & \multicolumn{2}{c}{RTO full corr} & \multicolumn{2}{c}{RTO partial corr}\\
		 $\mu_{C}$ & $\sigma_{C}$  & $\mu_{C}$ & $\sigma_{C}$  & $\mu_{C}$ & $\sigma_{C}$\\
		\cmidrule(r){1-2} \cmidrule(r){3-4} \cmidrule(r){5-6}
		 5.449 E5 & 3.076 E5 & 5.448 E5 & 3.066 E5 & 2.361 E5 & 9.708 E4\\
		\bottomrule
	\end{tabular}
	\label{example3_tab2}
\end{table*}

This example clearly shows that the nature of the distributed load has a 
significant effect on the RTO. Also from Table~\ref{example3_tab2}, it is 
possible to see that the compliance low-order statistics for a 2D bridge under 
a distributed load, emulated by a partially correlated random field, are smaller than 
those for a fully correlated field.


\section{Conclusions}
\label{concl_remaks}

In the present paper RTO problem has been formulated and solved 
by means of an optimization procedure which integrates a classical 
TO algorithm with a stochastic spectral expansion based on gPC.  
Monte Carlo simulation is used to verify the accuracy and efficiency of the proposed methodology. This approach 
is introduced to reduce the variability due to uncertain loadings applied to the mechanical 
structure of interest. The objective function of the robust problem is defined
as the weighted sum of the mean and standard deviation of the compliance, and it can be 
computed by considering a number of additional load cases.
This makes the RTO computationally tractable and accessible by any TO algorithm. 
Furthermore, the gPC is compatible with RTO for computing the statistical measures 
of the compliance. The numerical examples presented here show a substantial benefit 
and exhibits topology changes within their design domains compared with their 
deterministic counterpart. The optimal topology configurations confirm that the 
uncertainty parameters might change the deterministically obtained optimal topologies. 
The proposed methodology allows to obtain approximate outcomes with a much lower 
computational cost than that associated with Monte Carlo simulation, which makes it attractive, 
particularly in the context of structural topology optimization. Moreover, when using random load fields, the results 
show different topologies because the forces are correlated, i.e., each force depends on the 
other and therefore, their interactions with the structure have significant effects on the robust design.
The limitation of the gPC can be observed when a large number of random variables 
is used to parametrize the stochastic model, since in this case a substantial number of terms is
necessary to construct the expansion, and, consequently, the computational cost increases significantly 
with the dimension. This is often referred to as the curse of dimensionality, and it can be reduced 
using adaptive techniques such as the adaptive sparse grid.


\section*{Acknowledgments}


NC acknowledges the financial support from the Group of Technology in Computer Graphics (Tecgraf/PUC-Rio), Rio de Janeiro, Brazil. AP and IFMM acknowledge the financial support from the National Council for Scientific and Technological Development (CNPq) under projects 312280/2015-7 and 309708/2015-0, respectively.
AP and ACJr are thankful for the support from Carlos Chagas Filho Research Foundation of Rio de Janeiro State (FAPERJ) under grants E-26/203.189/2016, 
E-26/010.002.178/2015 and E-26/010.000.805/2018.
The information provided in this paper is the sole opinion of the authors and does not necessarily reflect the views of the sponsoring agencies.


\bibliographystyle{spbasic}      
\bibliography{bibliography}

\end{document}